\documentclass[useAMS,usenatbib]{mnras}

\usepackage{aas_macros}
\usepackage{graphicx}
\usepackage{amsmath}
\usepackage{amssymb}
\usepackage{enumerate}

\usepackage{siunitx}
\DeclareSIUnit\gauss{G}
\newcommand{\mg}[1]{\SI{#1}{\micro\gauss}}

\usepackage{xcolor}
\usepackage[thinc]{esdiff}

\title[Radio halo of NGC~5775]{CHANG-ES XXIII: Influence of a galactic wind in NGC~5775}
\author[G.~H.~Heald et al.]
{\parbox{\textwidth}{
G.~H. Heald,$^{1}$\thanks{E-mail: George.Heald@csiro.au}
V. Heesen,$^{2}$
S.~S. Sridhar,$^{3,4}$
R. Beck,$^{5}$
D.~J. Bomans,$^{6}$
M. Br\"uggen,$^{2}$
K.~T. Chy\.zy,$^{7}$
A. Damas-Segovia,$^{8}$
R.-J. Dettmar,$^{6}$
J. English,$^{9}$
R. Henriksen,$^{10}$
S. Ideguchi,$^{11}$
J. Irwin,$^{10}$
M. Krause,$^{5}$
J.-T. Li,$^{12}$
E.~J. Murphy,$^{13}$
B. Nikiel-Wroczy\'nski,$^{7}$
J. Piotrowska,$^{7}$
R.~J. Rand,$^{14}$
T.~Shimwell,$^{3,15}$
Y. Stein,$^{16}$
C.~J. Vargas,$^{17}$
Q.~D. Wang,$^{18}$
R.~J. van~Weeren,$^{15}$ and
T. Wiegert$^{10}$
}\vspace{0.4cm}\\
\parbox{\textwidth}{$^{1}$CSIRO, Space and Astronomy, PO Box 1130, Bentley, WA 6102, Australia\\
$^{2}$University of Hamburg, Hamburger Sternwarte, Gojenbergsweg 112, 21029 Hamburg, Germany\\
$^{3}$ASTRON, the Netherlands Institute for Radio Astronomy, Postbus 2, 7990 AA, Dwingeloo, The Netherlands\\
$^{4}$Kapteyn Astronomical Institute, University of Groningen, Landleven 12, 9747AD Groningen, The Netherlands\\
$^{5}$Max-Planck-Institut f\"ur Radioastronomie, Auf dem H\"ugel 69, 53121 Bonn, Germany\\
$^{6}$Ruhr University Bochum, Faculty of Physics and Astronomy, Astronomical Institute, 44780 Bochum, Germany\\
$^{7}$Astronomical Observatory, Jagiellonian University, ul. Orla 171, 30-244, Krak\'ow, Poland\\
$^{8}$Instituto de Astrof\'isica de Andaluc\'ia (CSIC), Glorieta de la Astronom\'ia, 18008 Granada, Spain\\
$^{9}$Department of Physics and Astronomy, University of Manitoba, Winnipeg, Manitoba, R3T 2N2, Canada\\
$^{10}$Dept. of Physics, Engeneering Physics, \& Astronomy, Queen’s University, Kingston, Ontario, Canada, K7L 3N6\\
$^{11}$Department of Astrophysics/IMAPP, Radboud University Nijmegen, PO Box 9010, 6500 GL Nijmegen, the Netherlands\\
$^{12}$Department of Astronomy, University of Michigan, 311 West Hall, 1085 S. University Ave., Ann Arbor, MI, U.S.A.\\
$^{13}$National Radio Astronomy Observatory, 520 Edgemont Road, Charlottesville, VA 22903, U.S.A.\\
$^{14}$Department of Physics and Astronomy, University of New Mexico, MSC07 4220, Albuquerque, NM, 87131, U.S.A.\\
$^{15}$Leiden Observatory, Leiden University, P.O. Box 9513, 2300 RA Leiden, The Netherlands\\
$^{16}$Observatoire astronomique de Strasbourg, Universit\'e de Strasbourg, CNRS, UMR 7550, 11 rue de l'Universit\'e, 67000 Strasbourg, France\\
$^{17}$Department of Astronomy and Steward Observatory, University of Arizona, 933 N Cherry Ave, Tucson, AZ, 85719 U.S.A.\\
$^{18}$Astronomy Department, University of Massachusetts, 710 N. Pleasant St., Amherst, MA 01003-9305, U.S.A.
}}

\begin{document}

\date{Accepted 2021 September 23. Received 2021 September 22; in original form 2020 August 7.}

\pagerange{\pageref{firstpage}--\pageref{lastpage}} \pubyear{2021}

\maketitle

\label{firstpage}

\begin{abstract}
We present new radio continuum images of the edge-on starburst galaxy NGC~5775, from LOFAR (140~MHz) and the Karl G. Jansky Very Large Array CHANG-ES survey (1500~MHz). We trace the non-thermal radio halo up to 13~kpc from the disc, measuring the non-thermal spectral index and estimating the total equipartition magnetic field strength ($\approx$\mg{13} in the disc and $\approx$\mg{7} above the plane). The radio halo has a similar extent at both frequencies, displays evidence for localized cosmic ray streaming coinciding with prominent H$\alpha$ filaments and vertical extensions of the regular magnetic field, and exhibits a boxy morphology especially at 140~MHz. In order to understand the nature of the disc-halo flow, we extend our previous model of cosmic ray propagation by implementing an iso-thermal wind with a tunable `flux tube' (approximately hyperboloidal) geometry. This updated model is successful in matching the vertical distribution of non-thermal radio emission, and the vertical steepening of the associated spectral index, in a consistent conceptual framework with few free parameters. Our new model provides the opportunity to estimate the mass outflow driven by the star formation process, and we find an implied rate of $\dot{M}\approx3$--$6\,\mathrm{M_{\sun}\,yr^{-1}}$ ($\approx40$--80 per cent of the star formation rate) if the escape velocity is reached, with substantial uncertainty arising from the poorly-understood distribution of ISM material entrained in the vertical flow. The wind may play a role in influencing the vertical gradient in rotational velocity.
\end{abstract}


\section{Introduction}
\label{sec:intro}

In the outskirts of galaxies, various processes key to their evolution take place. It is in these regions that gas accretes into the interstellar medium (ISM), where it can become fuel for star formation \citep[e.g.,][]{sancisi_etal_2008}. On the other hand, star-formation driven outflows can play an important role in expelling material and energy back into circumgalactic medium (CGM) or even in exceptional circumstances into the intergalactic medium (IGM). This is especially important in the case of starburst galaxies where the energetics are sufficient to accelerate matter out of the galaxy's potential well with significant consequences for their properties and evolution \citep{veilleux_etal_2005}, and in dwarf galaxies where the potential well is particularly shallow \citep{martin_1998,chyzy_etal_2016}.

Magnetic fields are an essential component of the physics in the outer regions of galaxies. The gas accretion process, seemingly required to fuel star formation \citep{sanchez_etal_2014,putman_2017}, may be facilitated by magnetic fields \citep{konz_etal_2002,galyardt_shelton_2016} and yet on the other hand may hamper condensation of clouds from the disc-halo interface \citep{gronnow_etal_2018}. The structure and energetics of outflows are also substantially impacted by the presence and features of the entrained magnetic fields \citep{heesen_etal_2011}. Generally speaking, the dynamo process is crucial for understanding how magnetic fields reached their current strength and structure in the local Universe \citep{beck_etal_2019}. Recent advances in numerical modelling reinforce the importance of the magnetized medium and the impact of resulting cosmic ray (CR) driven winds for galaxy evolution \citep[e.g.,][and references therein]{crocker_etal_2021}. In this context it has also been recognized that CR-driven galactic winds are important for the build-up of the CGM \citep{ji_etal_2020}.

A broad range of observational techniques have been brought to bear to probe the galaxy-IGM interface, and specifically the connection to the star formation process. The distribution and kinematics of extraplanar ISM gas are probed across the electromagnetic spectrum and particularly with emission line kinematics in the radio and optical bands \citep[e.g.,][]{putman_etal_2012}, while mapping of broadband continuum emission is a complementary technique that traces thermal Bremsstrahlung radiation, magnetic fields and cosmic rays 
in the galaxy-IGM interface \citep{condon_1992}. Studies of galaxies viewed at different inclination angles tend to be required in order to develop a holistic picture: face-on and intermediately-inclined galaxies are useful for directly associating extraplanar features with individual star forming regions, while edge-on galaxies are crucial for fully studying the vertical distribution.

Studies of the kinematics of extraplanar ISM material have revealed a common feature of galaxy gaseous haloes\footnote{In this paper, we use the word `halo' to refer to gas, dust, cosmic rays, and the magnetic field above and below the galaxy disc, not to be confused with stellar or dark matter haloes. Specifically, we call emission on larger scales, at vertical height $z>1$~kpc, halo emission, while the disc-halo interface is at $0.2<z<1$~kpc \citep[as defined by][]{irwin_etal_2012}.}: an overall decrease in rotation curve with increasing distance from the midplane \citep[e.g.,][]{sofue_etal_1992,rand_1997,swaters_etal_1997,zschaechner_etal_2015}. While this qualitative behavior is expected from three-dimensional disc-halo flow models such as the `galactic fountain' \citep{bregman_1980,norman_ikeuchi_1989}, it is clear that simple models incorporating only ballistic motion \citep{collins_rand_2002,fraternali_binney_2006} cannot reproduce the steepness of observed velocity gradients. Later models incorporating additional drag effects have proven more successful \citep[see, e.g.,][]{fraternali_2017} but a full understanding may still require additional effects to be understood, such as magnetic fields \citep{benjamin_2000,henriksen_irwin_2016}.

A particularly useful view of magnetized galaxy haloes is now becoming available at low radio frequencies ($\nu\lesssim350\,\mathrm{MHz}$). At these frequencies, the synchrotron and inverse-Compton (IC) energy losses suffered by CRs are relatively low, so magnetic fields can be traced far from the sites of CR acceleration \citep{hummel_etal_1991,carilli_etal_1992}. Low-frequency observations are powerful in combination with observations at higher radio frequencies, which provide a crucial benchmark for the distribution of radio spectral index, and where synchrotron polarization reveals the ordered magnetic field \citep[e.g.,][]{krause_etal_2020}.

Resolved studies of nearby galaxies at low radio frequency are now feasible thanks to the aperture array interferometers that have been constructed in recent years: the Murchison Widefield Array \citep[MWA;][]{tingay_etal_2013} and the LOw Frequency ARray \citep[LOFAR;][]{vanhaarlem_etal_2013}. LOFAR in particular is now providing access to the low frequency sky at high sensitivity and angular resolution, and all nearby galaxies in the northern sky will be imaged as part of the LOFAR Two-metre Sky Survey \citep[LoTSS;][]{shimwell_etal_2017}. The first radio halo investigated with LOFAR was that of NGC~891, where the scale heights of the non-thermal halo emission at 146~MHz were found to be larger than those at 1500~MHz by a factor of $1.7\pm0.3$ \citep{mulcahy_etal_2018}.

At higher radio frequencies, our view of the extended synchrotron properties of nearby galaxies has been dramatically improved with Continuum HAlos in Nearby Galaxies: an EVLA Survey \citep[CHANG-ES\footnote{\url{https://www.queensu.ca/changes/}};][]{irwin_etal_2012}. The survey was carried out with the Karl G. Jansky Very Large Array (VLA). CHANG-ES has been successful in probing the magnetic fields in and around galaxies \citep[see, e.g.,][]{krause_etal_2020}; a recent review of the survey and its primary outcomes is provided by \citet{irwin_etal_2019}.

To better understand the mechanisms at work within the CGM of galaxies, it is important to focus on individual targets that are known to drive strong disc-halo flows. In this paper, we focus on NGC~5775 (UGC~9579), an edge-on galaxy which is classified as a starburst \citep[see, e.g.,][]{tuellmann_etal_2006} featuring strong star formation activity that is not centrally concentrated but is instead widespread across the entire disc \citep[similar to another CHANG-ES sample galaxy, NGC~4666;][]{stein_etal_2019}. NGC~5775 has been well studied in the past, and is a prominent example of the class of objects with bright and filamentary extraplanar ISM features \citep{irwin_1994,collins_etal_2000,tuellmann_etal_2006}. The radio continuum halo has been studied several times previously, including by \citet{hummel_etal_1991} who first identified radio continuum emission extending above the star forming disc in this galaxy. Deeper observations were carried out by \citet{duric_etal_1998}, who mapped the radio continuum emission up to vertical heights $z=10$--$15\,\mathrm{kpc}$ and interpreted the spectral index distribution above the plane as arising from CR propagation and energy losses. \citet{irwin_etal_1999} put the radio halo in the context of a larger galaxy sample, arguing that CR diffusion alone was insufficient to describe the data, and tying the extraplanar emission to energy injection from star formation. \citet{tuellmann_etal_2000} revealed the structure of the ordered magnetic fields, identifying a strong vertical component associated with features seen in the diffuse ionized gas (DIG). \citet{soida_etal_2011} presented a deep multi-frequency polarization analysis of the three-dimensional magnetic field structure, concluding that both a galactic dynamo and the influence of a galactic wind were required to explain the field geometry.

The kinematics of the extraplanar gas in NGC~5775 have also been well studied. Long-slit spectroscopy revealed that the DIG at large vertical heights was rotating more slowly than the disc \citep{rand_2000,tuellmann_etal_2000}, but the lack of a complete three-dimensional view precluded full tilted-ring analysis. \citet{heald_etal_2006} returned to NGC~5775 with Fabry--Perot spectroscopy and found that the amplitude of the rotation curve decreases to higher vertical distances, at a rate of $7\,\mathrm{km\,s^{-1}\,kpc^{-1}}$ (corrected for the distance assumed in this paper). Most recently, \citet{boettcher_etal_2019} confirmed a vertical lag, and suggested that local values may exceed the global value from \citet{heald_etal_2006}, with uncertainty deriving from the lack of a three-dimensional kinematic model. Even on the basis of the relatively moderate lag reported by \citet{heald_etal_2006}, it is clear from their analysis of ballistic-model fountain orbits \citep{collins_rand_2002} that additional dynamical effects are required, as described above.

NGC~5775 has a close companion galaxy, NGC~5774. Signs of interaction are clearly visible through tidal bridges observed in multiple tracers including prominent features in {\sc H\,i} and radio continuum \citep{irwin_1994,duric_etal_1998}; a detailed discussion is provided by \citet{lee_etal_2001}. A smaller companion galaxy, IC~1070, may also be involved in the interaction. While the interaction is not strong enough to drive any substantial disturbance in the optical morphology of either NGC~5775 or NGC~5774, it is important to bear in mind that there may be impacts in the distribution and kinematics of the tenuous CGM many kpc from the star-forming disc.

In this paper we build on the previous radio continuum observations by adding new low-frequency imaging that is ideally matched in sensitivity and angular resolution to the data products collected by the CHANG-ES project. Our aims are to enhance our view of the extended synchrotron halo of NGC~5775, and thereby to clarify the distribution and strength of magnetic fields, details of the CR propagation, and their influence on the structure and kinematics of the galaxy.

The global properties of NGC~5775 are summarized in Table~\ref{table:n5775properties}. We have adopted the distance employed for NGC~5775 by the CHANG-ES survey, $D=28.9\,\mathrm{Mpc}$, based on the Hubble flow \citep{irwin_etal_2012}. Throughout the paper, spectral index is defined such that $S\propto\nu^\alpha$.

\begin{table*}
\caption{Properties of NGC~5775.}
\label{table:n5775properties}
\begin{tabular}{llc}
\hline
Parameter & Value & Reference \\
\hline
Right Ascension (J2000.0) & $14^\mathrm{h}53^\mathrm{m}57\fs 57$ & 1 \\ 
Declination (J2000.0) & $+03\degr 32\arcmin 40\farcs 1$ & 1 \\ 
Adopted distance & 28.9 Mpc & 2 \\ 
Angular scale & 7.14~arcsec\,kpc$^{-1}$ & 2 \\ 
\hline
Diameter of the stellar disc ($D_{25}$) & $4.17~{\rm arcmin} = 35.0\,\mathrm{kpc}$ & 3 \\
Diameter of the star forming (H$\alpha$) disc & $3.58~{\rm arcmin} = 30.1\,\mathrm{kpc}$ & 4 \\
Diameter of the radio continuum disc & $4.7~{\rm arcmin} = 39.5\,\mathrm{kpc}$ & 4 \\
Adopted (H$\alpha$+IR) star formation rate (SFR)$^\dagger$ & $7.56\pm0.65\,\mathrm{M_{\sun}\,yr^{-1}}$ & 5 \\
SFR surface density & $(9.41\pm0.81)\times10^{-3}\,\mathrm{M_{\sun}\,yr^{-1}\,kpc^{-2}}$ & 5 \\
Alternative (1.4~GHz) SFR$^\dagger$ & $22.9\,\mathrm{M_{\sun}\,yr^{-1}}$ & 4 \\
\hline
Inclination & $85\fdg 8$ & 1 \\ 
Position Angle & $145\fdg 7$ & 1 \\
Maximum midplane $v_\mathrm{rot}$ & $198\,\mathrm{km\,s^{-1}}$ & 1 \\
Vertical rotation gradient (${\rm d}v_{\rm rot}/{\rm d}z$) & $1\,\mathrm{km\,s^{-1}\,arcsec^{-1}}=7\,\mathrm{km\,s^{-1}\,kpc^{-1}}$ & 6 \\
\hline
\multicolumn{3}{l}{References: (1) \citet{irwin_1994}; (2) \citet{irwin_etal_2012}; (3) \citet{rc3};}\\
\multicolumn{3}{l}{(4) this work; (5) \citet{vargas_etal_2019}; (6) \citet{heald_etal_2006}.}\\
\multicolumn{3}{l}{$^\dagger$ From the \citet{murphy_etal_2011} calibrations, assuming a Kroupa IMF between $0.1$--$100\,\mathrm{M_\odot}$.}\\
\end{tabular}
\end{table*}

This paper is organized as follows. We describe the observations and data reduction in Section~\ref{section:data}. The properties of the radio halo in NGC~5775 are described in Section~\ref{section:properties}, and new cosmic ray (CR) modeling is presented in Section~\ref{sec:transport}. We discuss the data and CR model in a broader context in Section~\ref{section:discussion}. We conclude the paper in Section~\ref{section:conclusions}, including some thoughts on future studies of radio haloes.

\section{Observations and data reduction}\label{section:data}

\subsection{LOFAR}\label{subsection:lofar}

LOFAR data were collected for project LC1\_046 over the course of two nights, 6--7 and 7--8 May 2014, for a total on-source integration time of 9.6 hours. We chose to observe in two short sessions on subsequent nights in order to achieve a long combined integration, while also avoiding times of low source elevation where both sensitivity and $(u,v)$ coverage are degraded. The frequency range covered by this observation is $117.6$--$189.8$~MHz. Visibilities were recorded in all four linear polarizations (XX,YY,XY,YX) with 1~s time resolution, and 64 channels per 195.3~kHz subband (200~MHz clock); the recorded channel width is 3.1~kHz. We made use of all Dutch LOFAR stations in the HBA\_DUAL\_INNER configuration, which maximises the similarity of the primary beam size between the differently-sized core and remote stations. International stations were not used.

The primary calibrator source 3C~295 was observed before and after the primary target on each night, with the same frequency coverage as the main target. A secondary calibrator (UGC~9799) was observed simultaneously with NGC~5775 (in this case with reduced frequency coverage), but we do not make use of that additional beam in this paper.

We performed initial calibration of the visibility data using the standard LOFAR imaging pipeline \citep{heald_2018}. We used the Averaging Pipeline to flag and average the data. After flagging of edge channels in each subband, and RFI flagging using {\tt aoflagger} \citep{offringa_etal_2012} with the standard HBA settings (as provided for example as part of {\sc prefactor}\footnote{\url{https://github.com/lofar-astron/prefactor}}), the pipeline averaged the data to an intermediate frequency resolution of 48.8~kHz and time resolution of 2~s. Demixing \citep{vdtol_etal_2007} was not performed. 

The subsequent data reduction procedure followed the `facet calibration' approach presented by \citet{vanweeren_etal_2016} and \citet{williams_etal_2016}. Calibration was carried out in two phases: a direction-independent step followed by a direction-dependent step. In the direction-independent step, we derived and corrected the data for amplitude gains, clock offsets, offsets between the X and Y dipoles, and ionospheric rotation measure. In this step, we also performed phase calibration using a $25$-$\rm arcsec$ model of the sky derived from The GMRT Sky Survey \citep[TGSS;][]{intema_2017}.

Using the direction-independent calibrated data, we generated a total sky model of all the sources within the field of view. We then subtracted the sky model from the visibility data and divided the field into multiple facets such that each facet has at least one point-like facet calibrator source brighter than 0.4~Jy. For this dataset, we used 25 facets to cover a field spanning a diameter of about 10 degrees. For each facet, we added the model of the facet calibrator source and performed four iterations of amplitude and phase self-calibration. The new calibration solutions were used to image the full associated facet, and the corresponding model was subtracted from the visibilities. This process was followed sequentially for all facets. The final facet to be imaged was the one containing NGC~5775, using calibration solutions from an adjoining facet to avoid difficulties with self-calibrating the target galaxy itself. The resulting visibility data have a frequency resolution of 488.3~kHz and a time resolution of 10~s. 

The central facet containing NGC\,5775 was ultimately reimaged together with the CHANG-ES data (Section~\ref{subsection:changes}) as described in Section~\ref{subsection:finalimages}. 

\subsection{VLA}\label{subsection:changes}

VLA data were collected for the CHANG-ES survey (project 10C-119) on several dates as summarised in Table~\ref{table:changes}. Data reduction proceeded as described by \citet{irwin_etal_2012} and \citet{wiegert_etal_2015}; we do not repeat the details here. In brief, the data were flagged and calibrated in the {\it Common Astronomy Software Applications} \citep[{\sc casa};][]{mcmullin_etal_2007} package. The absolute flux density scale was set using 3C~286 on the Perley--Butler 2010 frequency scale \citep{perley_and_butler_2013}. Flagging was performed on the basis of visual inspection. Flagging and calibration were repeated iteratively until satisfactory results were achieved.

\begin{table*}
\caption{Summary of CHANG-ES L-band observations of NGC~5775.}
\label{table:changes}
\begin{tabular}{llll}
\hline
VLA configuration      & B                    & C                   & D                    \\
\hline
Observing date(s)      & 5 Apr 2011           & 30 Mar 2012         & 30 Dec 2011          \\
Integration time (min) & 116                  & 40                  & 18                   \\
Frequency (MHz)$^a$        & $1247$--$1503$,         & $1247$--$1503$,        & $1247$--$1503$,         \\
\,                     & $1647$--$1903$          & $1647$--$1903$         & $1647$--$1903$          \\
Bandwidth (MHz)        & 512                  & 512                 & 512                  \\
Flux calibrator        & 3C286                & 3C286               & 3C286                \\
Zero-pol calibrator    & OQ208                & OQ208               & OQ208                \\
Phase calibrator       & J1445+0958           & J1445+0958          & J1445+0958           \\
Min/max $(u,v)$ (m)         & 174.2/10883.3        & 58.4/3385.7         & 25.8/994.6           \\
Min/max $(u,v)$ ($\lambda$) & 724.0/69027.1        & 242.6/21473.8       & 107.4/6308.3         \\
\hline
\multicolumn{4}{l}{$^a$ The gap in frequency coverage was chosen to avoid strong RFI.}
\end{tabular}
\end{table*}

Starting from the calibrated data, the B-, C-, and D-configuration \emph{L}-band data were imaged together using {\sc wsclean} \citep{offringa_etal_2014} with weighting and tapering tuned to match the LOFAR imaging, as described in Section~\ref{subsection:finalimages}.

\subsection{Final images}\label{subsection:finalimages}

\subsubsection{Total intensity imaging}

\begin{figure*}
    \includegraphics[width=\hsize]{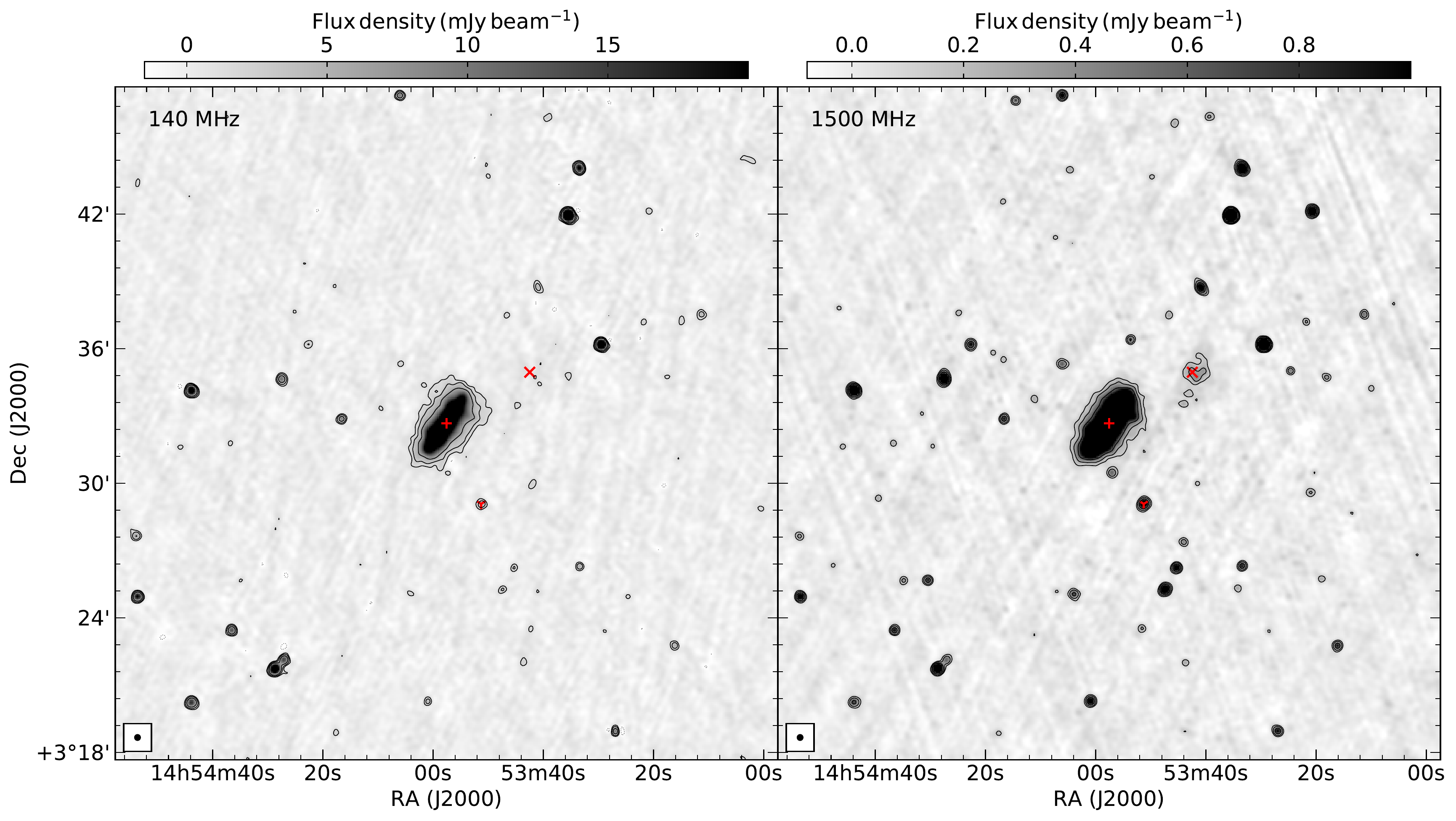}
    \caption{Radio continuum images of the NGC~5775 field. Both are represented in grayscale and with contours to illustrate the low surface brightness structure. For the LOFAR image at 140~MHz (left panel), black contours begin at 1.5~mJy~beam$^{-1}$ and increase by factors of 2. For the CHANG-ES image at 1500~MHz (right panel), black contours begin at 0.15~mJy~beam$^{-1}$ increasing by factors of 2. In both cases, light gray contours correspond to the negative of the first black contour levels (i.e., $-1.5$ and $-0.15$~mJy~beam$^{-1}$, respectively). The positions of NGC~5775, NGC~5774, and IC~1070 are shown with a red plus, cross, and Y-shape, respectively. The synthesized beam size is displayed in the bottom-left corner of each panel. Note that the full LOFAR field of view is substantially larger than the region displayed here.}
    \label{fig:widefield}
\end{figure*}

The final imaging steps for both the LOFAR and CHANG-ES data sets were performed with tuned parameters in order to result in consistent image resolution, as well as closely matched sensitivity to relevant angular scales. In this common imaging step, we excluded data beyond the maximum and minimum common baseline lengths as expressed in units of the observing wavelength, $39~\mathrm{k}\lambda$ and $110~\lambda$, respectively. We used {\sc wsclean} to produce all of the images, using multiscale multifrequency {\sc clean} \citep{offringa_smirnov_2017}. For both the LOFAR and CHANG-ES data sets, we used {\sc clean} scales ranging from 0 (point source) to approximately the angular size of NGC~5775. The output image from each data set is a single map with broad bandwidth formed via multifrequency synthesis (MF), although internally the {\sc clean} algorithm operated on 8 bandwidth segments, accounting for spectral structure within the broad LOFAR and VLA bandwidths. In each case, an initial image was formed using shallow {\sc clean} ($10\times$ the anticipated noise level). A mask was formed by performing source finding with {\sc pybdsf} \citep{mohan_rafferty_2015} on that initial image, and then a final image was produced by applying the {\sc clean} algorithm using the mask and with a threshold corresponding to the expected noise level. The consistent inner to outer $(u,v)$ range listed in Table~\ref{table:images} was selected, and imaging weights were chosen to provide consistent resolution without applying $(u,v)$-plane tapering. The final images are summarised in Table~\ref{table:images}. As summarised in the Table, minor additional smoothing was applied in the image plane in order to produce images with exactly matched resolution. The smoothing was performed in {\sc miriad} \citep{sault_etal_1995} and preserved the flux scale.

As mentioned in Section~\ref{sec:intro}, a radio continuum bridge connects NGC~5775 to its companion NGC~5774 \citep[see, e.g.,][]{duric_etal_1998}. We detect this feature using our new observations (see Section~\ref{sss:advection_velocity}), but it is best reproduced with lower-resolution images generated with a robust parameter closer to natural weighting and that emphasize larger angular scales. We focus in this paper on the angular resolution that optimizes our ability to constrain the cosmic ray propagation mechanism (Section~\ref{sec:transport}). We will return to the radio continuum bridge in more detail in a followup paper (English et al., {\it in prep}).

\begin{table}
    \caption{Summary of NGC~5775 imaging.}
    \begin{tabular}{ll}
        \hline
        LOFAR & \, \\
        Reference frequency & 140~MHz \\
        $(u,v)$ range ($\lambda$) & $110$--$10\,000$ \\
        Briggs weighting & $-0.3$ \\
        Imaging pixel size & $1.5$~arcsec \\
        Unconvolved beam size & $17.8\times 16.1$~arcsec$^2$\, (PA=$70\degr$) \\
        Final resolution & $18.5$~arcsec\, (2.2 kpc) \\
        Image noise & $440\,\umu\mathrm{Jy\,beam^{-1}}$ \\
        \hline
        CHANG-ES & \, \\
        Reference frequency & 1500~MHz \\
        $(u,v)$ range ($\lambda$) & $110$--$10\,000$ \\
        Briggs weighting & $+1.5$ \\
        Imaging pixel size & 4.5~arcsec \\
        Unconvolved beam size & $18.2\times 16.0$~arcsec$^2$\, (PA=$172\degr$) \\
        Final resolution & 18.5~arcsec\, (2.2 kpc) \\
        Image noise & $24\,\umu\mathrm{Jy\,beam^{-1}}$ \\
        \hline
    \end{tabular}
    \label{table:images}
\end{table}

Following direction-dependent calibration of the LOFAR data as described in Section~\ref{subsection:lofar}, the astrometric scale is not necessarily aligned with the CHANG-ES image. Before proceeding, we performed source finding on the two final images at consistent resolution using {\sc bane} and {\sc aegean} \citep{hancock_etal_2018}, and cross matched the resulting catalogs to identify a mean position angular offset with a magnitude of $(\Delta\alpha,\Delta\delta)=(0.19\arcsec,1.84\arcsec)$ in Right Ascension and Declination. An astrometric correction was applied to the final LOFAR image to correct for this offset. After the correction, the source finding procedure was repeated, demonstrating that the relative astrometry is matched to within $\lesssim0.05~\rm arcsec$.

The total intensity images are presented in Figure~\ref{fig:widefield}. The morphology of the radio halo is discussed in Section~\ref{section:properties}.

\begin{figure*}
    \centering
    \includegraphics[width=\hsize]{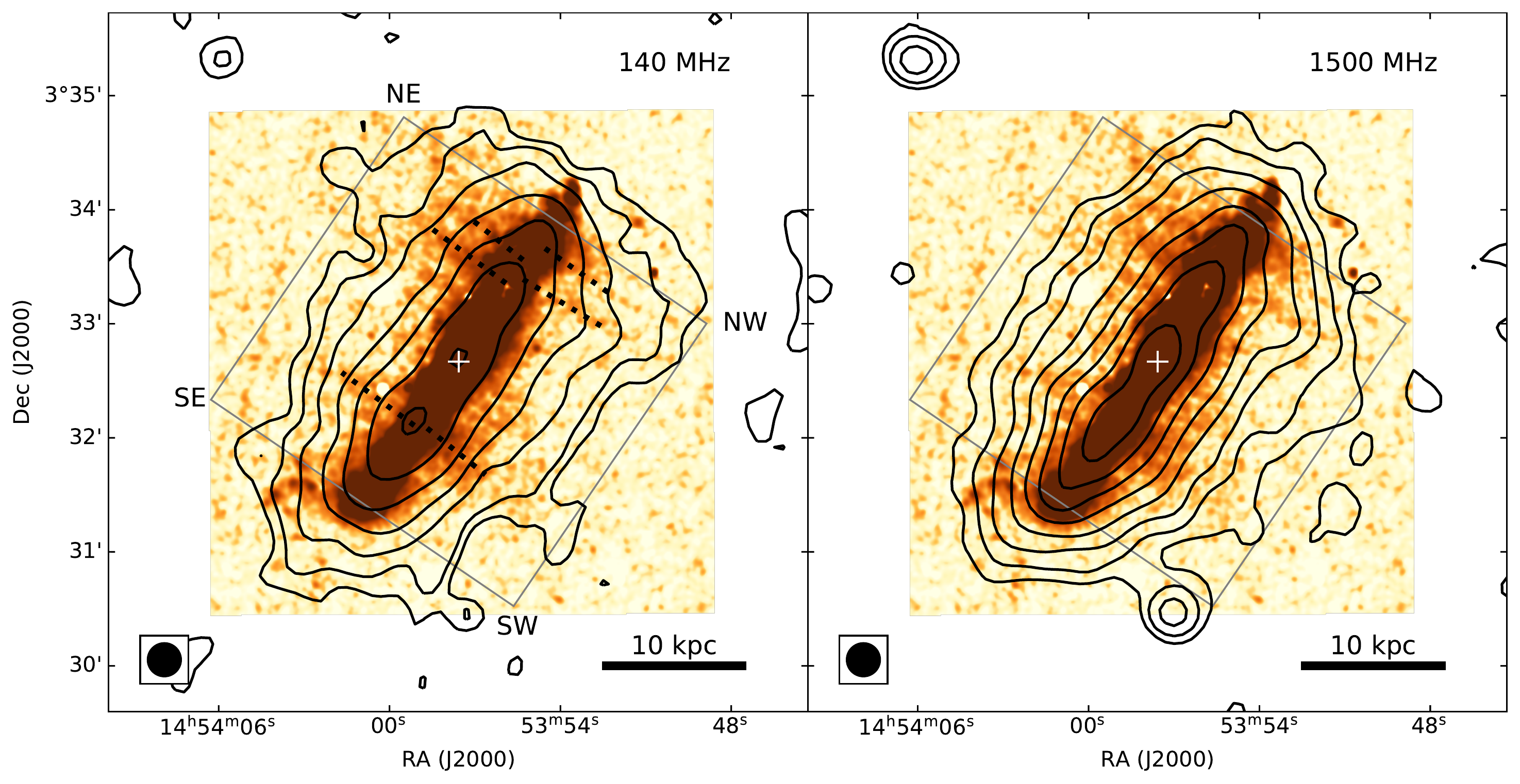}
    \caption{Contours of non-thermal radio continuum emission from NGC~5775. The background colormap is the continuum-subtracted H$\alpha$ image from \citet{collins_etal_2000}, smoothed with a $2~\rm arcsec$ Gaussian kernel and presented with a log stretch to emphasize faint filamentary features in the thick disc region. The image is overlaid with black contours corresponding to non-thermal radio emission derived from the 140~MHz LOFAR image (left panel) and 1500~MHz CHANG-ES image (right panel), as described in the text. Contour levels start at $1\,\mathrm{mJy\,beam^{-1}}$ for LOFAR, at $0.1\,\mathrm{mJy\,beam^{-1}}$ for CHANG-ES, and in both cases increase by powers of two. The beam size ($18.5~\rm arcsec$) of the radio images is shown in the bottom left corner of each panel. In the left panel, locations of prominent H$\alpha$ filaments are displayed with dotted lines to guide the eye. Also in the left panel we have annotated the NE, NW, SE, and SW quadrants as they are referred to throughout the paper. The position of the centre of NGC~5775 is indicated with a white plus. The gray box indicates the region considered for the analysis of the vertical distribution in Section~\ref{subsec:vertdist}.}
    \label{fig:halpha}
\end{figure*}

\subsubsection{Thermal subtraction}\label{subsubsec:thermal}

The radio images include a non-negligible contribution from thermal Bremsstrahlung radiation, which makes it difficult to isolate and interpret the synchrotron emission and thus constrain the properties of the magnetic fields and cosmic rays. The thermal contribution was estimated through the use of the H$\alpha$ map from \cite{collins_etal_2000} and a {\it Spitzer} MIPS 24-\si{\micro}m map from the NASA/IPAC Infrared Science Archive's Spitzer Heritage Archive (SHA)\footnote{\url{http://irsa.ipac.caltech.edu/applications/Spitzer/SHA/}} \citep{rieke_2004}. We quantified the thermal contribution to the total intensity radio continuum emission using the following equation from \cite{hunt_2004}:

\begin{multline} \label{eq:extinction}
\left( \frac{F_\nu}{\mathrm{mJy}} \right) = 1.16 \left( 1 + \frac{n(\mathrm{He}^+)}{n(\mathrm{H}^+)} \right) \left( \frac{T}{10^4~\mathrm{K}} \right)^{0.617} \\ \times \left( \frac{\nu}{\mathrm{GHz}} \right)^{-0.1} \left( \frac{F_{H\alpha\mathrm{,corr}}}{10^{-12}~\mathrm{erg~cm}^{-2}~\mathrm{s}^{-1}} \right).
\end{multline}

In equation~\eqref{eq:extinction}, $F_\nu$ is the estimated radio flux density due to thermal emission at frequency $\nu$ and $F_{H\alpha\mathrm{,corr}}$ is the extinction-corrected H$\alpha$ flux. An extinction correction was applied to the H$\alpha$ image following the recipe described by \cite{calzetti_2007}. $T$ is the electron temperature within the emitting region and is assumed to be $10^4$~K. Following \cite{martin_1997}, we have also assumed that the ratio of the number density of ionized helium to that of ionized hydrogen $n(\mathrm{He}^+)/n(\mathrm{H}^+)$ is 0.087.

Using this method, we determine that the mean thermal fraction at 1500 (140) MHz is about 15 (4) per cent, with higher values up to 57 (23) per cent found at the locations of star-forming regions in the disc. That the thermal contribution reaches such high values in isolated regions supports the conclusion drawn by \citet{irwin_etal_2019} regarding localized areas of relatively shallow spectral index in NGC~5775, and more broadly is consistent with measurements of hundreds of individual star forming regions across a large sample of nearby galaxies \citep{linden_etal_2020}. The thermal contribution is relatively high even at LOFAR frequencies, consistent with the fact that NGC~5775 hosts active and widespread star formation activity.

We note that an alternative method for estimating the thermal contribution has recently been published by \citet{vargas_etal_2018}. It is possible that through the analysis employed here, and specifically the use of the \citet{calzetti_2007} extinction correction, we have underestimated the thermal contribution within the disc by a factor of approximately 1.36 \citep{vargas_etal_2018}. However, all such estimates carry substantial uncertainty for edge-on galaxies, and it is unlikely that the systematic difference between these two procedures will change any of the main conclusions in the present paper. In particular, our estimate of the non-thermal contribution will have most effect in the midplane, but will be largely inconsequential in the halo where we focus most of our attention.

The estimated thermal contribution to our radio continuum images was subtracted to produce maps of the non-thermal radio continuum emission. The resulting distribution at 140 and 1500~MHz is shown in comparison to the sensitive H$\alpha$ image from \citet{collins_etal_2000} in Figure~\ref{fig:halpha}.

\section{Properties of the radio halo}\label{section:properties}

In this section, we describe the features of the non-thermal radio halo of NGC~5775 as seen with our new images, highlighting aspects that build on features already known from previous work.

\subsection{Vertical distribution}\label{subsec:vertdist}

The most striking feature of the radio halo in NGC~5775 is the highly extended distribution of diffuse emission away from the midplane, up to a characteristic height of about 13~kpc from the star-forming disc on both sides (at the surface brightness indicated by the lowest contours in Figure~\ref{fig:halpha}). The radio halo was known to be extended from previous studies \citep{duric_etal_1998,soida_etal_2011,krause_etal_2018}; here we have traced the halo to approximately the same vertical extent, and crucially added the low frequency image from LOFAR. The very wide frequency span provided by the two images separated by a decade in frequency is essential to constrain the cosmic ray propagation and magnetic field properties in the halo. In these new images, the vertical extent is nearly as large as the radial size. This effect is similar to what has been seen in other galaxies in the gaseous and non-thermal distributions \citep[e.g.,][]{oosterloo_etal_2007,stein_etal_2019}, and is understood to be indicative of a distribution driven by vertical motions.

\begin{figure}
    \includegraphics[width=\hsize]{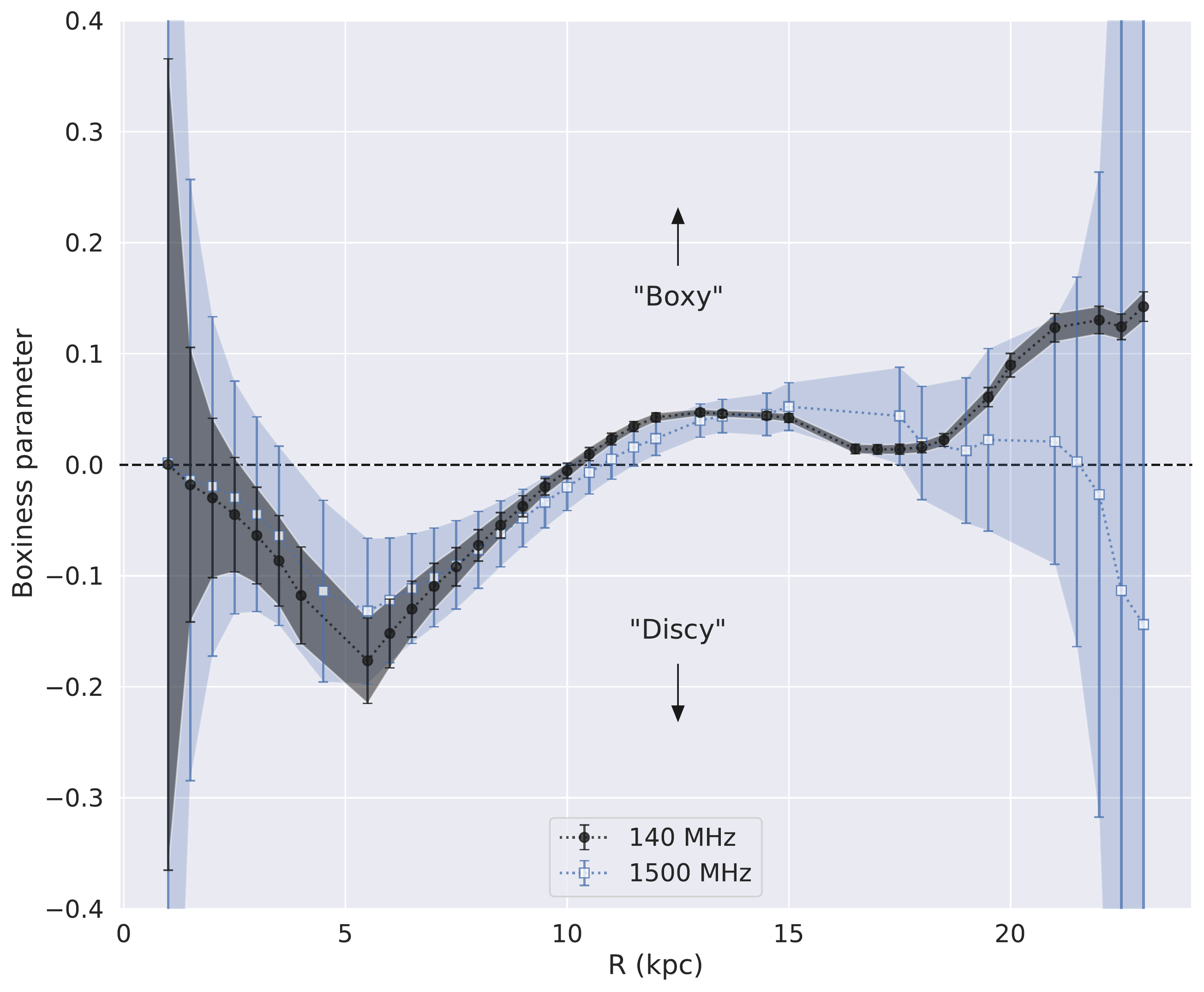}
    \caption{Boxiness parameter for the non-thermal halo of NGC~5775, calculated as described in the text. Positive values indicate a boxy morphology, while negative values indicate a discy morphology. Errorbars are indicative of the sample mean at that radius.}
    \label{fig:boxy}
\end{figure}

Another striking feature is that the morphology is apparently boxier at low frequencies than at GHz frequencies. We have quantified this through a technique typically used in optical astronomy to parameterize the shapes of elliptical galaxies. As described for example by \citet{bender_etal_1988}, a harmonic decomposition around azimuthal angle of the radial deviation between an isophote and its best-fitting ellipse reveals through the amplitude of the $\cos(4\theta)$ term an indication of `boxiness' and `disciness'. Typically, this value is normalized by the characteristic radius and the local surface brightness gradient \citep[e.g.,][]{ciambur_2015}, yielding what we call here the `boxiness parameter'. Due to the normalizsation by the local gradient, the boxiness parameter is negative for discy isophotes, and positive for boxy isophotes. We have calculated this metric using the {\sc python} {\tt photutils.isophote} module \citep{photutils}, and the result is visualised in Figure~\ref{fig:boxy}. At both frequencies, the radio morphology has a discy character at intermediate radii, while the low-frequency morphology is clearly boxy at the largest radii ($R\gtrsim20$~kpc). This is consistent with the more general statement that the low-frequency radiation dominates most strongly at large vertical extent and the outer radii. Interestingly, despite the generally X-shaped ordered magnetic field \citep[][and see Figure~\ref{fig:spix}]{soida_etal_2011,krause_etal_2020}, the radial extent of the radio halo is essentially constant all the way up to $z\approx13\,\mathrm{kpc}$. We return to this in Section~\ref{section:discussion}.

Inspection of the contours in Figure~\ref{fig:halpha} shows that the morphology at lower frequency is also more diffuse (i.e., smeared out) than at higher frequency. This is a qualitative way to recognise that the overall spectral index distribution exhibits a steepening to higher $z$. This is quantified and examined in more detail in Section~\ref{subsec:spix}.

We have fitted exponential scale heights to the vertical distribution of the non-thermal emission at both frequencies, in each of the quadrants separately. The scale height fitting includes the beam size as a component of the procedure, following the approach described by \citet{krause_etal_2018} and assuming $i=90\degr$. Specifically, we extracted vertical profiles with 1~kpc sampling ($\approx2.5$ samples per synthesized beam) from each quadrant, for measured values inside the star forming radius which we take to be $R<90$~arcsec $(12.6\,\mathrm{kpc})$, and for heights $|z|<96$~arcsec $(13.5\,\mathrm{kpc})$. This region is illustrated in Figure~\ref{fig:halpha}. For each quadrant, we fitted a function describing the convolution of an intrinsic exponential vertical profile
\begin{equation}
I(z) = I_0\,e^{-z/h_\nu}
\end{equation}
with the Gaussian beam of the images
\begin{equation}
G(z) = \frac{1}{\sqrt{2\pi\sigma^2}}\,e^{-z^2/2\sigma^2},
\end{equation}
where $h_\nu$ is the vertical scale height at frequency $\nu$, and $\sigma$ is related to the synthesized beam (FWHM $18.5$~arcsec) through $\mathrm{FWHM}=2\sqrt{2\ln{2}}\sigma$. 
The convolution takes the form \citep[see][]{mueller_etal_2017}:
\begin{equation}
\begin{split}
I_\mathrm{conv}(z) & = \frac{I_0}{2}\,e^{-z^2/2\sigma^2}\\
\times\,\Biggl\{ & \exp \left[ \left( \frac{\sigma^2-zh_\nu}{\sqrt{2}\sigma h_\nu}\right)^2 \right]\,\mathrm{erfc} \left(\frac{\sigma^2-zh_\nu}{\sqrt{2}\sigma h_\nu}\right)\\
+ & \exp \left[ \left( \frac{\sigma^2+zh_\nu}{\sqrt{2}\sigma h_\nu}\right)^2 \right]\,\mathrm{erfc} \left(\frac{\sigma^2+zh_\nu}{\sqrt{2}\sigma h_\nu}\right) \Biggr\},
\end{split}
\label{eqn:conv}
\end{equation}
where $\mathrm{erfc}(x)$ is the complementary error function. For completeness, we note that this treatment does not account for radial variation of the radio surface brightness. However, because NGC~5775 is viewed from an edge-on perspective we are unable to reliably establish and incorporate the radial surface brightness profile in the likely presence of expected non-axisymmetric morphological features and localised surface brightness fluctuations. We therefore consider the integrated vertical structure in a similar fashion to \citet{krause_etal_2018}.

The functional form in equation~\eqref{eqn:conv} was fit to the data for each quadrant using the \texttt{curve\_fit} function in the \textsc{python} \texttt{scipy.optimize} module. We used the Levenberg--Marquardt algorithm, and provided uncertainties for each data point based on the standard deviation of measured values within each vertical profile bin. Formal errors for the fitted scale heights were obtained from the covariance matrix returned by the \texttt{curve\_fit} function.

\begin{figure*}
    \includegraphics[width=\hsize]{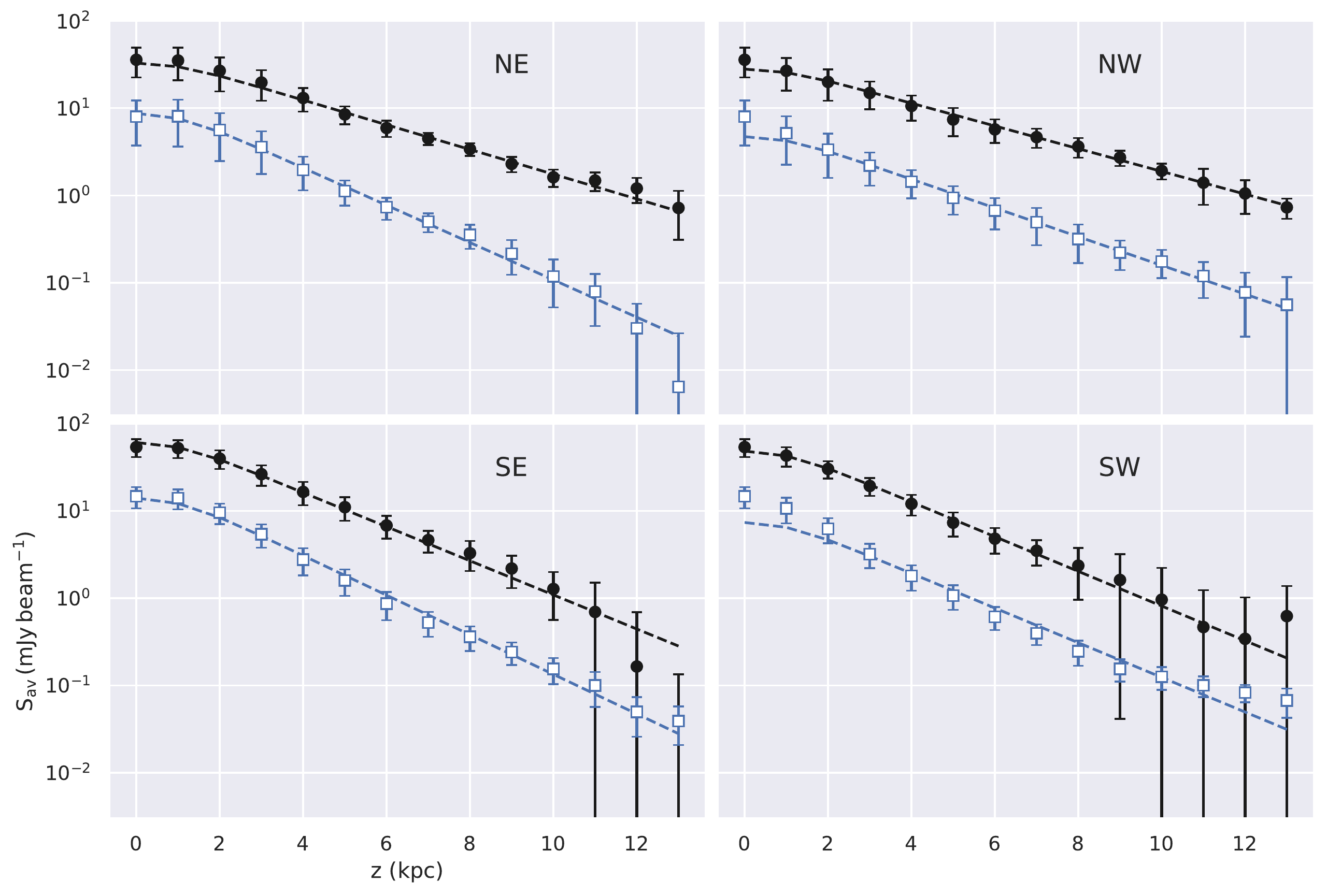}
    \caption{Vertical distribution of LOFAR (filled black circles) and CHANG-ES (open blue squares) non-thermal emission, with fitted exponential distributions (including resolution effects corresponding to the $18.5\arcsec$ beam) overlaid with dashed lines. Fitted scale height values are provided in Table~\ref{tab:scaleheights}.}
    \label{fig:vertdist}
\end{figure*}

The vertical profiles for the LOFAR and CHANG-ES images in each quadrant, along with the fitted function (equation~\ref{eqn:conv}), are displayed in Figure~\ref{fig:vertdist}. The corresponding fitted scale height ($h_\nu$) values and their associated formal errors are provided in Table~\ref{tab:scaleheights}. The scale height at 140~MHz is for all quadrants larger than or equal to that at 1500~MHz, with an average scale height ratio of $1.2\pm0.3$. This weak frequency dependence is somewhat smaller than the value derived for NGC~891 by \citet{mulcahy_etal_2018}, $1.7\pm0.3$, and indicates that synchrotron losses are small. Interestingly, there is no indication that we require more than a single exponential component for each quadrant. For this same galaxy \citet{krause_etal_2018} required separate `disc' and `halo' components. In part, this may suggest that the thermal contribution is nearly exclusive to the thin disc and thus after the subtraction of the thermal component as described in Section~\ref{subsubsec:thermal}, the thick disc component dominates so that we only require a single scale height to reproduce the profiles. In the NW and SW quadrants, the fitted exponential distribution slightly underestimates the measured values in the midplane ($z\lesssim$~1~kpc), which may be a consequence of a small underestimate of the thermal contribution as mentioned in Section~\ref{subsubsec:thermal}. We also note that \citet{krause_etal_2018} used images with an angular resolution of 11~arcsec, and were therefore better able to resolve the non-thermal thin disc component. Their analysis also used seven vertical strips across the full radial extent of NGC~5775, whereas we use only two. Regardless of the relative importance of these factors, our non-thermal images are plainly dominated by emission from the thick disc.

\begin{table}
    \caption{Best-fitting exponential scale heights (in kpc) for the non-thermal thick disc, their frequency dependence ($h\propto\nu^\zeta$), and best-fitting exponential scale height (in kpc) of equipartition magnetic field strength.}
    \begin{tabular}{lcccc}
        \hline
        \, & $h_\mathrm{140\,MHz}$ & $h_\mathrm{1500\,MHz}$ & $\zeta$ & $h_\mathrm{Beq}$ \\
        \hline
        NE & $3.07\pm0.24$ & $2.03\pm0.17$ & $-0.17\pm0.05$ & $13.68\pm1.79$ \\
        NW & $3.34\pm0.26$ & $2.64\pm0.28$ & $-0.10\pm0.06$ & $16.87\pm2.09$ \\
        SE & $2.22\pm0.16$ & $1.92\pm0.10$ & $-0.06\pm0.04$ & $12.45\pm1.14$ \\
        SW & $2.19\pm0.21$ & $2.19\pm0.14$ & $+0.00\pm0.05$ & $16.12\pm1.15$ \\
        \hline
    \end{tabular}
    \label{tab:scaleheights}
\end{table}

\subsection{Spectral index and equipartition magnetic field}\label{subsec:spix}

\begin{figure*}
    \centering
    \includegraphics[width=0.48\hsize]{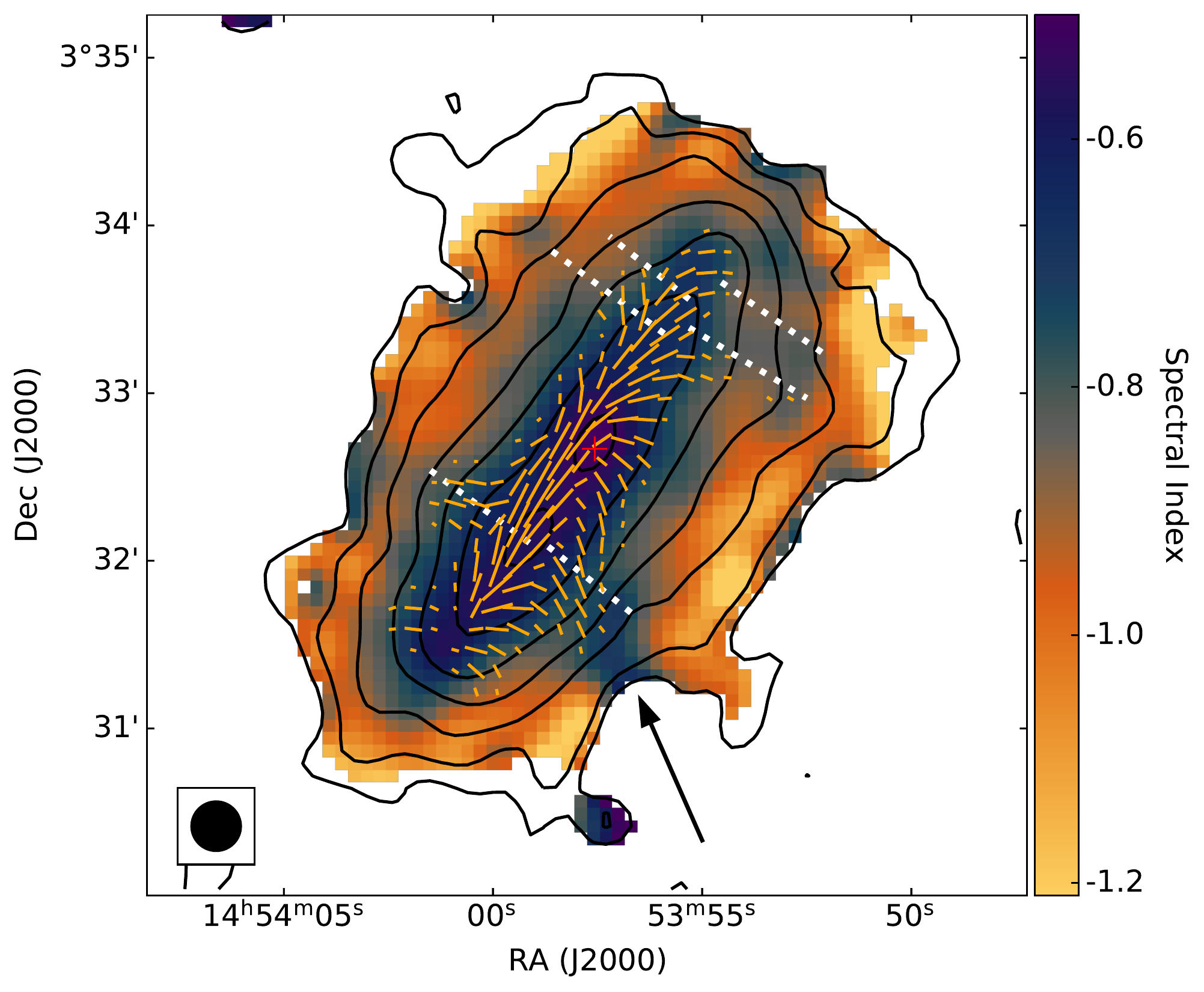}\hfill
    \includegraphics[width=0.48\hsize]{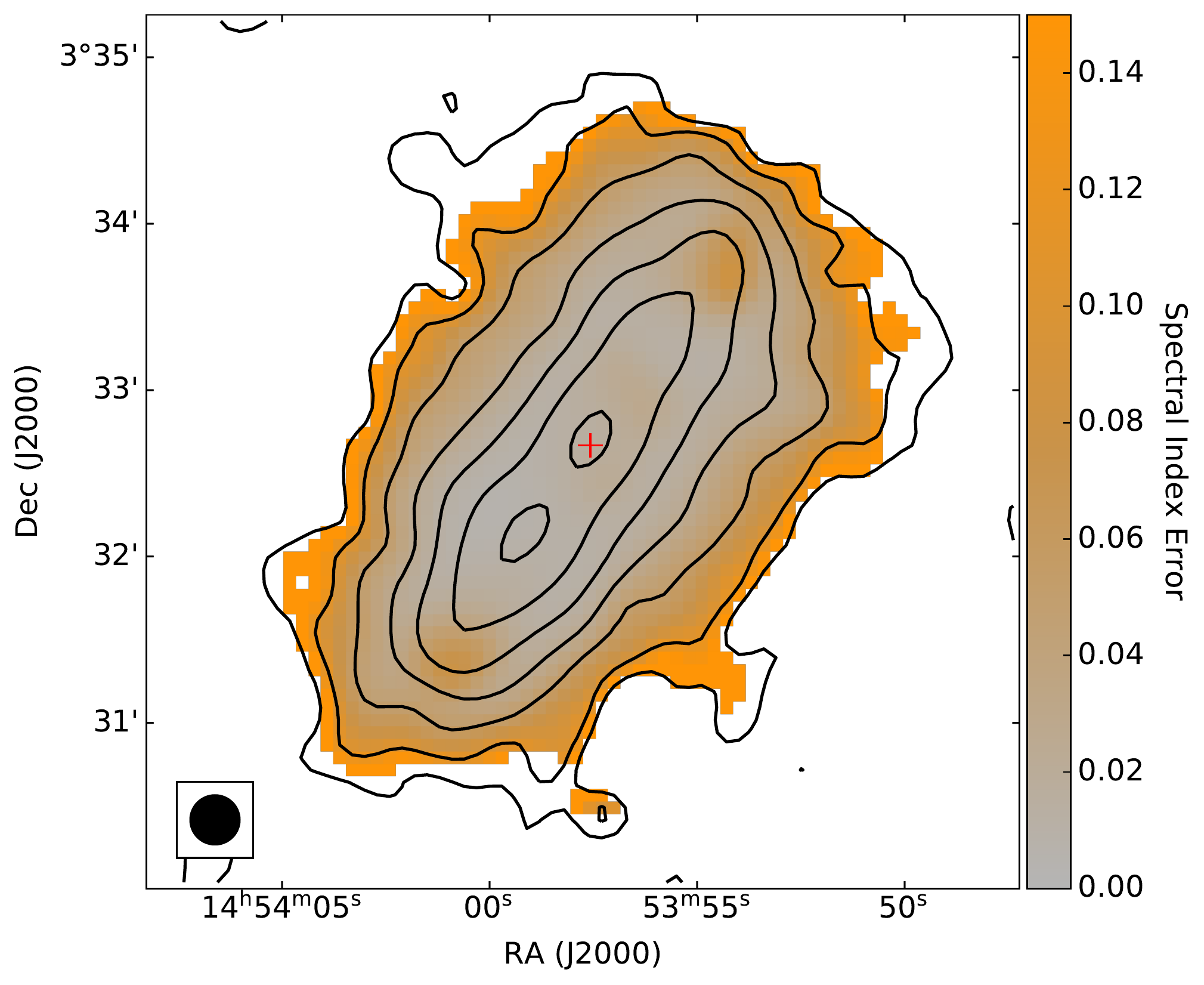}
    \caption{Left: Non-thermal spectral index map constructed from the thermal-corrected LOFAR and CHANG-ES maps. Locations of prominent H$\alpha$ filaments are displayed with white dotted lines, as in Figure~\ref{fig:halpha}. The ordered magnetic field as determined by \citet{krause_etal_2020} is shown with orange vectors. The channel of shallow spectral index associated with the filament in the SW quadrant, discussed in the text (Section~\ref{subsec:spix}), is indicated with an arrow. Right: Spectral index error estimated as described in the text. In both panels, the black contours show the distribution of 140~MHz radio continuum emission (starting at 1~mJy~beam$^{-1}$ and increasing by powers of two), the centre of the galaxy is marked with a red plus, and the beam size is shown with a black circle at the bottom left corner.}
    \label{fig:spix}
\end{figure*}

With the thermal contribution to the total intensity removed, we can now determine the spectral index associated with the synchrotron radiation. We estimated the spectral index through
\begin{equation}
\frac{S_\mathrm{1500,nth}}{S_\mathrm{140,nth}} = \left( \frac{\mathrm{1500\,MHz}}{\mathrm{140\,MHz}} \right)^\alpha,
\end{equation}
where $S_\mathrm{1500,nth}$ and $S_\mathrm{140,nth}$ are the non-thermal contributions to the broadband CHANG-ES and LOFAR maps, which we take to have characteristic radio frequencies of 1500~MHz and 140~MHz, respectively. The spectral index map was computed for pixels where the LOFAR and CHANG-ES maps exceed $1\,\mathrm{mJy\,beam^{-1}}$ and $0.1\,\mathrm{Jy\,beam^{-1}}$, respectively. The spectral index distribution is shown in Figure~\ref{fig:spix}.

Uncertainties in the non-thermal spectral indices were determined by considering the noise levels in the non-thermal maps at both frequencies, as well as estimated errors in the thermal subtraction procedure. The latter was achieved through standard error propagation using Equation~\ref{eq:extinction} and assuming fractional errors of 10\% in each of the temperature, extinction-corrected H$\alpha$ flux, and ratio of ionized helium to ionized hydrogen parameters \citep[see also][]{vargas_etal_2018}, ultimately leading to an estimated 14\% fractional error on the thermal estimate. The resulting spectral index errors are also shown in Figure~\ref{fig:spix}. Note that these errors do not account for possible spectral curvature, or for a possible systematic underestimate of the thermal contribution (see Section~\ref{subsubsec:thermal}) which would apply nearly equally at both frequencies and lead to only a small impact on the spectral index.

The most prominent feature of the spectral index distribution is the clear steepening from the disc to the halo (see also Figure~\ref{fig:vertprofs}). This effect has been seen previously in this galaxy \citep[e.g.,][]{soida_etal_2011}. The primary cause is energy losses suffered by CR$e^-$ as they propagate away from the regions in the disc where supernova remnants associated with ongoing star formation accelerate the particles to their injection energy. This process has been studied in detail in other galaxies with particular progress in recent years \citep[e.g.,][]{heesen_etal_2018,vargas_etal_2018}. We address this in detail for NGC~5775 in Section~\ref{sec:transport}.

The locations of some of the brighter and more prominent H$\alpha$ filaments, and the distribution and orientation of the ordered magnetic fields \citep[from][]{krause_etal_2020}, are also shown in Figure~\ref{fig:spix}. Remarkably, the locations of the H$\alpha$ filaments appear to coincide with regions of the halo where high-$z$ spectral indices are shallower, suggesting that the CR$e^-$ in those regions have suffered less energy loss than in other regions. Moreover, the ordered magnetic field shows clear vertical extensions at the same locations, with the field orientation apparently aligned with the `channels' of shallower spectral index. CRs can stream along magnetic field lines and hence locally have higher transport speeds \citep{wiener_17a}. This then could explain the flatter gradient of spectral indices, since ageing of the CR$e^-$ is locally diminished. This is particularly clear for the filament in the SW quadrant, where the radio emission in the halo retains a spectral index $\alpha\approx-0.7$ up to $z\approx10\,\mathrm{kpc}$, while the surrounding emission rapidly steepens to typical spectral index values $\alpha\lesssim-1$. If attributed to an under-estimation and -subtraction of the local thermal radio continuum, our estimate in this region from the procedure described in Section~\ref{subsubsec:thermal} would need to be incorrect by a factor of $\sim5$--$10$, which we deem to be unlikely. As noted in Section~\ref{subsubsec:thermal}, our estimate of the thermal contribution is possibly underestimated in the disc by a much smaller factor of about 1.36 \citep{vargas_etal_2018}, which would lead to a typical spectral index error of only $\Delta\alpha\approx0.02$. We return to the observed correspondence between H$\alpha$ morphology, spectral index distribution, and ordered magnetic field orientation in Section~\ref{section:discussion}.

\begin{figure}
    \centering
    \includegraphics[width=\hsize]{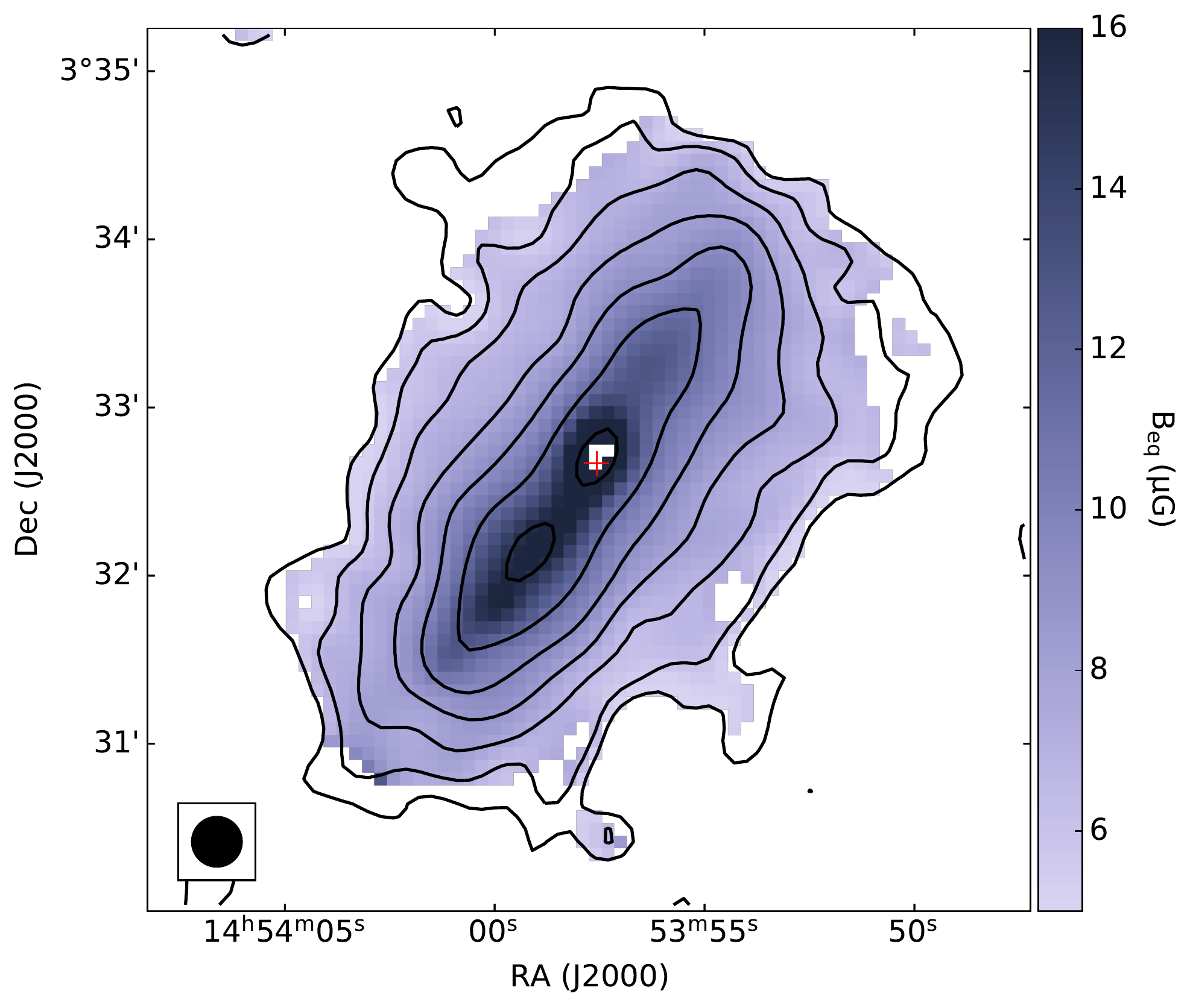}
    \caption{Equipartition magnetic field strength in \si{\micro}G. The black contours show the distribution of 140~MHz radio continuum emission (starting at 1~mJy~beam$^{-1}$ and increasing by powers of two), the centre of the galaxy is marked with a red plus, and the beam size is shown with a black circle at the bottom left corner.}
    \label{fig:beq}
\end{figure}

\begin{figure*}
    \centering
    \includegraphics[width=0.7\hsize]{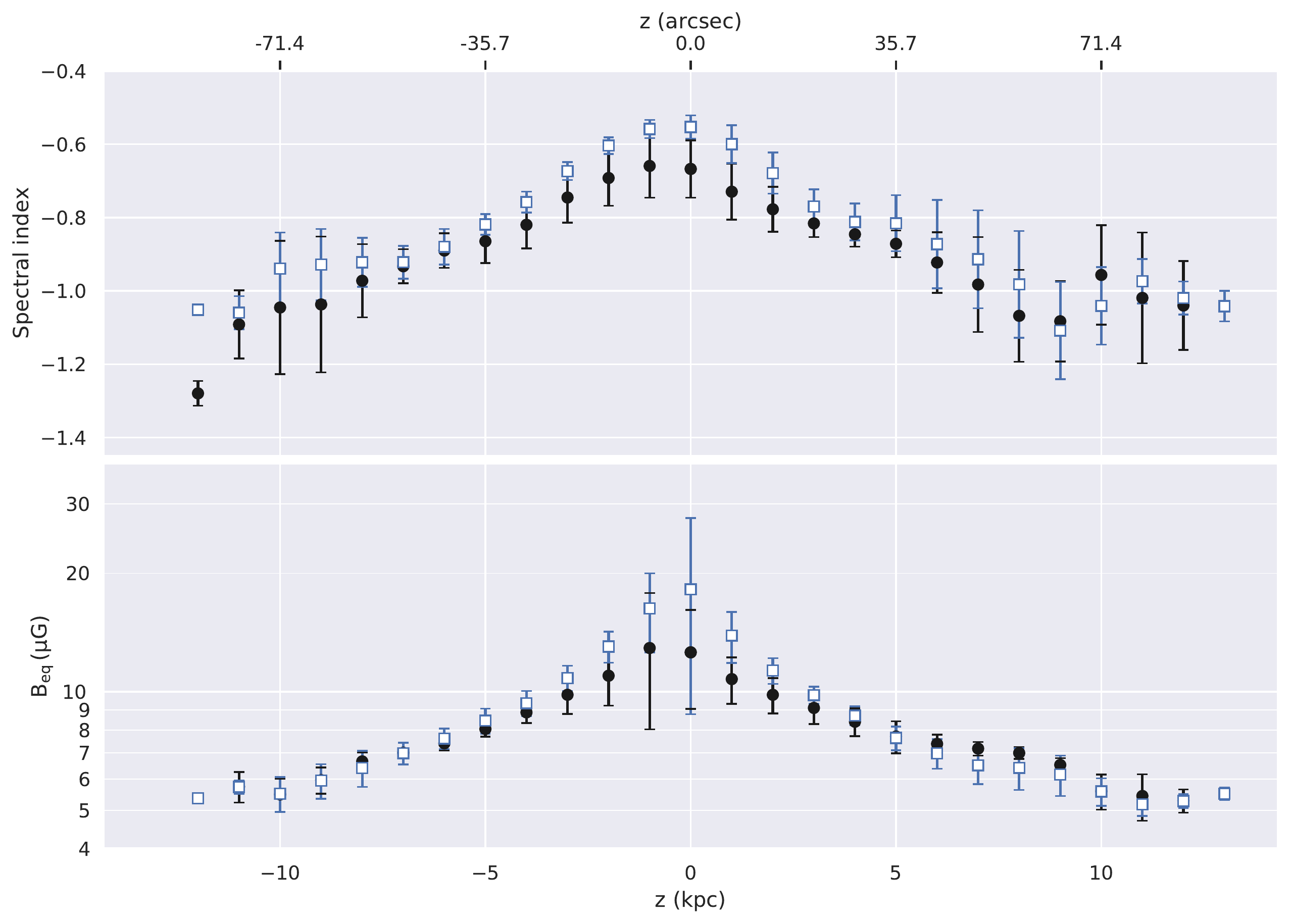}
    \caption{Vertical profiles of spectral index (top) and equipartition magnetic field strength (bottom). Profiles are plotted separately for the north side of the disc (filled black circles) and the south side (open blue squares). Errorbars indicate the standard deviation of values within the corresponding vertical bin.}
    \label{fig:vertprofs}
\end{figure*}

On the basis of the spectral index distribution, we can now compute the total magnetic field strength averaged along the line of sight as implied by the assumption of energy equipartition with the cosmic rays \citep[dominated by protons;][]{beck_krause_2005}. Equipartition is not guaranteed and indeed is not expected to hold below the kpc scale \citep[see][]{seta_beck_2019}, but larger-scale deviations in magnetic field strength as large as an order of magnitude from the equipartition value can be excluded on the basis of the typical non-thermal morphology of galaxies \citep{duric_1990}. We require estimates of the local pathlength ($l$) through the synchrotron-emitting medium, and $K_0$, the ratio of proton to electron number densities per unit energy interval (for our purposes, at energies of a few GeV). To approximate $l$, we take the local projected distance through an edge-on cylindrical volume extending above and below the disc, with a radius corresponding to the maximum radial extent of the radio continuum detected from NGC~5775 ($20\,\mathrm{kpc}$; see Table~\ref{table:n5775properties}). For the proton-to-electron ratio, we take the canonical value of $K_0=100$. We neglected pixels where the non-thermal spectral index was steeper than $-1.2$, to avoid overestimating the magnetic field strength in regions where strong synchrotron loss has occurred and the extrapolation to the low-energy portion of the CR spectrum is expected to be incorrect with the \citet{beck_krause_2005} method. The resulting distribution of equipartition magnetic field strength ($B_\mathrm{eq}$) is shown in Figure~\ref{fig:beq}. The average magnetic field strength is about \mg{8} across the entire galaxy, with relatively high values above \mg{20} in the central region. The span of total magnetic field strength that we have derived for NGC~5775 is consistent with the higher end of the range that has been established through radio continuum observations of a variety of spiral galaxies over the years \citep[see, e.g., the compilation presented by][their Table 3]{beck_etal_2019}. The derived midplane field strength is about 50 per cent higher in the southern disc than in the northern disc (\mg{19} and \mg{13}, respectively), and everywhere drops off slowly with increasing distance from the disc, as shown in Figure~\ref{fig:vertprofs}. To broadly characterize the vertical profile, we fitted exponential scale heights for $B_\mathrm{eq}(z)$ following the same method as was used for the synchrotron emission; these values are presented in Table~\ref{tab:scaleheights}.

\section{Galactic wind}
\label{sec:transport}

\subsection{Physical picture}
\label{subsec:Physical_picture}

Our new radio images have revealed the distribution of non-thermal radio emission in the thick disc of NGC~5775, with clearly distinct vertical structure across the decade in frequency span. We now seek to model the propagation of CRs in the thick disc, with a particular aim to better understand the dynamics of the disc--halo flow driven by the disc-wide starburst in NGC~5775. We assume that the thermal gas and the CRs can be modelled as polytropic gases with adiabatic indices of $\gamma_{\rm g}=5/3$ and $\gamma_{\rm CR}=4/3$, respectively. This kind of setup has been investigated in the literature with 1D wind models that take both the CRs and the thermal gas into account \citep{breitschwerdt_etal_1991,everett_etal_2008,everett_et_al_2010, samui_et_al_2010, recchia_etal_2016}. These papers have shown that one can formulate a `wind equation' that includes pressure terms from both the thermal and the CR gas. The composite sound speed, including both pressure contributions, is then nearly constant. CRs are vital in this setup since they prevent the wind from cooling adiabatically as they are transported faster than the wind fluid. The radio spectral index analysis in Section~\ref{subsec:spix} suggests that the CRs are transported faster along the vertical magnetic field lines at the locations of the prominent H$\alpha$ filaments, which hints at either CR anisotropic diffusion or localised streaming.    

In this work, we sidestep the question of the importance of CR streaming and/or (anisotropic) diffusion for the necessary energy transport, a topic which is currently extensively debated in the literature \citep[e.g.][]{wiener_17a,farber_18a,chan_et_al_2019}. Instead, we solve the Euler (momentum conservation) equation, where we assume for simplicity that the speed of sound is constant. Assuming pure CR advection, we are able to test whether typical wind velocity profiles are consistent with the radio data. We note that the details of CR transport are important to investigate whether CRs are indeed able to \textit{drive} the wind, and to study their relative importance in comparison to thermal and radiation pressure \citep{yu_et_al_2020}. This is largely left to future work, but we show in Appendix~\ref{as:cosmic_ray_driven_wind} that the physical properties of our iso-thermal wind model can indeed be reproduced by a fully self-consistent model that explicitly takes into account CR streaming as a driving agent.

If we conservatively assume that the pressure to launch the wind stems from the CRs alone, and $\dot E_{\rm CR}$ is the CR luminosity, $\dot M$ is the mass flux, and $v$ is the (asymptotic) wind velocity, then energy conservation requires $\dot E_{\rm CR} = 1/2 \epsilon \dot M v^2$, where we introduced the parameter $\epsilon$ representing the efficiency of entraining thermal mass into the wind. In order to ensure that the wind is energetically feasible, we require $\epsilon\leq 1$. For $\epsilon=1$, the CRs lose all their energy due to adiabatic expansion and transfer it to the gas. This is an extreme case, but as we will show in Section~\ref{section:mass_loss_rate}, the entrainment efficiency is of order unity, so only a moderate correction is needed in order to ensure energy conservation. We search for wind solutions with given energy densities and pressures of the hot, thermal X-ray emitting gas \citep{li_etal_2008} and of the total CRs as obtained from energy equipartition with the magnetic field. The required advection speeds in the halo are at least $v\approx 300~\rm km\,s^{-1}$, so that with canonical diffusion coefficients of $D\approx 10^{28}~\rm cm^2\,s^{-1}$ the length-scale where diffusion dominates over advection is $D/v\lesssim 0.1~\rm kpc$. This is much smaller than the size of the of the halo ($\sim 10~\rm kpc$) and of the resolution provided by our radio data, hence we neglect diffusion and use pure advection to describe the CR transport.

We make use of the SPectral INdex Numerical Analysis of K(c)osmic-ray Electron Radio-emission ({\sc spinnaker}\footnote{\url{https://github.com/vheesen/Spinnaker}}) package, as described by \citet{heesen_etal_2016,heesen_etal_2018}. {\sc spinnaker} is a 1-D cosmic-ray transport code that numerically solves equations for pure advection and diffusion, and calculates synthetic non-thermal radio continuum profiles for direct comparison with observational quantities. We assume a steady state where the CR$e^-$ are injected in the galactic midplane and are advected in a vertical direction, while losing energy due to synchrotron and inverse Compton radiation. Other CR particles are not included in the model because it is the electrons that dominate the synchrotron radiation from galaxies at these frequencies \citep[e.g.,][]{condon_1992}. We experimented first with matching the vertical profiles for NGC~5775 with a constant wind velocity and assuming an exponential magnetic field. However, we found a model with an accelerating wind to be equally suitable in terms of fit quality, and with the substantial added benefit of better fulfilling the energy equipartition condition (see Section~\ref{ss:comp_prev_models} for more details). An accelerating advection speed had already been initially and successfully explored as an approximation to a wind profile in the context of NGC~891 and NGC~3556 \citep[][respectively]{schmidt_etal_2019,miskolczi_etal_2019}. In the case of NGC~891, for example, the best-fitting advection flow was an accelerated galactic wind with midplane velocity $\approx150~\mathrm{km\,s^{-1}}$, reaching the escape velocity at a height of $9$--$17$~kpc, depending on location in the galaxy. In this paper, we further develop this model of a wind for the case of NGC~5775, implementing the wind parameters in a conceptual framework that consistently establishes vertical variation in key quantities with few free parameters.

\subsection{Main assumptions}
\label{ss:main_assumptions}

The main assumptions of our simplified galactic wind model are as follows:
\begin{enumerate}[(i)]
    \item the geometry is a tuneable `flux tube' (approximately hyperboloidal);
    \item the wind is iso-thermal; and
    \item the gravitational acceleration decreases exponentially in the vertical direction.
\end{enumerate}
The first assumption (i) ensures that the wind expands laterally with height above the plane, so that the magnetic field strength will decrease due to flux conservation. Note that that even though we use the flux tube approximation, we are dealing with a galaxy-scale outflow. Hence, we use a single flux tube with an initial radius of $r_0=7~\rm kpc$, which is approximately half the star-forming disc radius. Assumption (ii) means that the composite sound speed, of the gas and the CRs, is constant. Winds start subsonic, and exceed the sound speed at the so-called critical point to become supersonic. Because the sound speed is constant, the wind can only go through the critical point if the gravitational potential decreases. This is the case for the superposition of a disc-like and dark matter halo gravitational potential, which can be well approximated by an exponential function (assumption (iii)). The latter assumption neglects the influence of the companion galaxy NGC~5774. While we do see a bridge of radio emission between the two galaxies (as discussed in Section~\ref{sec:intro} and \ref{subsection:finalimages}), the overall influence on the radio halo seems to be weak, as there are no noticeable asymmetries in either the radio continuum emission or the radio spectral index (see Table~\ref{tab:scaleheights}). We will address the properties of the radio bridge in a future paper.

The primary limitation of our model is the assumption of a constant sound speed, which is actually unphysical since the adiabatic cooling of the cosmic rays and the thermal gas reduces their temperature. In a more realistic model, the temperature drops by a factor of a few within the extent of our radio halo, and the sound speed is reduced accordingly. This then reduces the ability of the gas and the cosmic rays to accelerate the wind further. Although it has not yet been directly tested with observational data, our indicative self-consistent model (see Appendix~\ref{as:cosmic_ray_driven_wind}) shows that in such a framework, the acceleration of the wind is indeed considerably reduced in the halo. While we attempt to correct for the inclusion of the unphysical source of energy by adjusting the entrainment efficiency to reassert energy conservation, our approach is by design partly phenomenological and thus offers limited conclusions.

\subsection{Iso-thermal wind model}
\label{ss:wind_model}

\begin{figure}
    \includegraphics[width=\hsize]{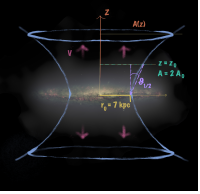}
    \caption{A cartoon sketch illustrating the overall geometry of the flux tube model employed here. The cross-sectional area $A(z)$ increases with height following Eqn.~\ref{eq:flux_tube}. The starting radius $r_0$, advection speed $v$, half opening angle $\theta_{1/2}$, and meaning of the characteristic height $z=z_0$ are also displayed. The background image is the total continuum halo from the VLA C-configuration only \citep[see][]{irwin_etal_2019_dr3}, overlaid with an optical image of NGC~5775 constructed from NASA/ESA {\it Hubble Space Telescope} data (F658N and F625W) obtained from the Hubble Legacy Archive.}
    \label{fig:cartoon}
\end{figure}

We assume that the CRs are advected in the flow of magnetised plasma. This flow is directed vertically and is expanding adiabatically. We assume the following functional term for the cross-sectional area (parallel to the disc):
\begin{equation}
    A(z) = A_0 \left [1 + \left (\frac{z}{z_0}\right )^\beta \right ].
    \label{eq:flux_tube}
\end{equation}
This form describes an expanding flow, which has been used previously in semi-analytic 1D wind models (see beginning of Section~\ref{sec:transport}). This particular choice eases the comparison with the aforementioned models, and provides a unifying picture within which we can approximate the vertical variations in the key quantities as we describe below. If $\beta=2$ then the model has a hyperboloidal shape, initially cylindrical near the plane and opening up asymptotically to the constant opening angle that would define the corresponding (bi-)conical form. The cross-sectional area then defines also the radius of the outflow via $A=\upi r^2$. The model geometry is illustrated in Figure~\ref{fig:cartoon}.

We also require an equation that governs the magnetic field strength:
\begin{equation}
    B = B_0 \left (\frac{r}{r_0}\right )^{-1} \left ( \frac{v}{v_0}\right )^{-1}, 
\end{equation}
where $B_0$ is the magnetic field strength in the galactic midplane, and $r_0$ and $v_0$ are the midplane flow radius and advection speed, respectively. This is the expected behaviour for radial and toroidal magnetic field components in an axisymmetric, accelerating, quasi-1D flow \citep{baum_etal_1997}. This prescription also ensures that the magnetic field is approximately in energy equipartition with the CRs, as $B^2\propto r^{-2}v^{-2}$ (see Section~\ref{sss:magnetic_field_strength} for further discussion). We tested that the magnetic field falls off in a similar way as the flux conservation would demand with $B\propto r^{-2}$ for the vertical component.

The next conditions that need to be fulfilled are described by the continuity equation and energy conservation. The continuity equation is:
\begin{equation}
    \rho v A = \rm constant,
    \label{eq:continuity}
\end{equation}
where $v$ is the advection speed and $\rho$ is the gas density. The flow of the plasma is governed by the Euler equation:
\begin{equation}
    \rho v \diff{v}{z} = -\diff{P}{z}  - g\rho, 
    \label{eq:euler}
\end{equation}
where $P=P_{\rm g}+P_{\rm CR}$ is the combined gas and CR pressure, and $g$ is the gravitational acceleration. Here, we assume that the wind is driven by the thermal gas in the hot phase and the CRs together. We now assume that the composite sound speed \citep{breitschwerdt_etal_1991},
\begin{equation}
    v_c^2 = \diff{(P_{\rm g}+P_{\rm CR})}{\rho},
\end{equation}
is constant. Because we do not  explicitly include CR streaming and diffusion, the composite sound speed is simply:
\begin{equation}
    v_c^2 = \frac{\gamma_{\rm g} P_{\rm g} + \gamma_{\rm CR} P_{\rm CR}}{\rho},
    \label{eq:composite_sound_speed}
\end{equation}
where $\gamma_g$ and $\gamma_{\rm CR}$ are the adiabatic indices of the thermal gas and CRs, respectively. This is the reason we refer to this wind model as iso-thermal even though we include the CRs as a non-thermal component. The aforementioned 1D wind models show that this is a good approximation within 20~kpc distance from the galactic midplane, beyond our detection threshold for the radio halo. Actually, the composite sound speed increases slightly in the halo, which is a consequence of a wind solution with a nearly constant gravitational acceleration \citep{mao_ostriker_2018}. Hence, our wind speeds are on the conservative (lower) side. Then we can write the Euler equation in the following way:
\begin{equation}
    \rho v \diff{v}{z} = -v_{\rm c}^2 \diff{\rho}{z} -g\rho.
\end{equation}
This equation contains only the velocity $v$ and the density $\rho$, where the latter can be eliminated using the continuity equation~\eqref{eq:continuity}:
\begin{equation}
    \left ( v - \frac{v_{\rm c}^2}{v} \right ) \diff{v}{z} = \frac{\beta v_{\rm c}^2z^{\beta-1}}{z_0^\beta}\left [1+\left(\frac{z}{z_0}\right )^\beta \right ]^{-1} - g.
\end{equation}
This `wind equation' has a critical point at $z=z_{\rm c}$ where $v=v_{\rm c}$ and ${\rm d}v/{\rm d}z$ is undefined.  Because the composite sound speed is constant, the gravitational acceleration needs to fall off with height in order to pass through the critical point. We have chosen an exponential fall-off with $g=g_0\exp(-z/h_{\rm grav})$, where $h_{\rm grav}$ is the gravitational scale height. Here $g_0=v_{\rm rot}^2/(2r_0)$ is the maximum gravitational acceleration which is found near the disc. We assumed $h_{\rm grav}= 4r_0 = 28~\rm kpc$, which is a bit smaller than the diameter of the star-forming disc as viewed in the H$\alpha$ image (see Table~\ref{table:n5775properties}). This scaling is modelled in such a way that it describes the gravitational acceleration in the halo of the Milky Way at the solar radius (Appendix~\ref{a:wind_models}).

Integrating the wind equation leads to the conservation of energy, equivalent to the Bernoulli principle:
\begin{eqnarray}
    \left ( \frac{v}{v_{\rm c}}\right )^2 - \log\left ( \frac{v}{v_{\rm c}} \right )^2 & = & 2 \log\left [ 1+\left (\frac{z}{z_0}\right )^\beta \right ] \nonumber \\ & + & \frac{v_{\rm rot}^2h_{\rm grav}}{v_{\rm c}^2r_0}\exp\left ( -\frac{z}{h_{\rm grav}} \right )\nonumber \\ & + & C.
    \label{eq:wind_solution}
\end{eqnarray}

The integration constant $C$ is chosen in such a way that the right-hand side ($rhs$) of equation~(\ref{eq:wind_solution}) fulfills $rhs=1$ at the critical point. This is the case if:
\begin{eqnarray}
    C & = & 1 - 2\log\left [ 1+\left (\frac{z_{\rm c}}{z_0}\right )^\beta\right ] \nonumber \\
  & - &   \frac{v_{\rm rot}^2h_{\rm grav}}{v_{\rm c}^2r_0}\exp\left ( -\frac{z_{\rm c}}{h_{\rm grav}} \right ).
\end{eqnarray}
Note that the right-hand side of equation~(\ref{eq:wind_solution}) has a minimum at the critical point. The expanding flow leads to a rising potential (first term of the $rhs$); on the other hand the decreasing gravitational potential combines with the expanding flow to create a minimum. We solve equation~\eqref{eq:wind_solution} numerically after searching for the minimum of the $rhs$, after which we can calculate the integration constant $C$. The derivation of the wind equation and its solution is explained in more detail in Appendix~\ref{a:wind_models}.

\begin{figure*}
\centering
\includegraphics[width=\hsize]{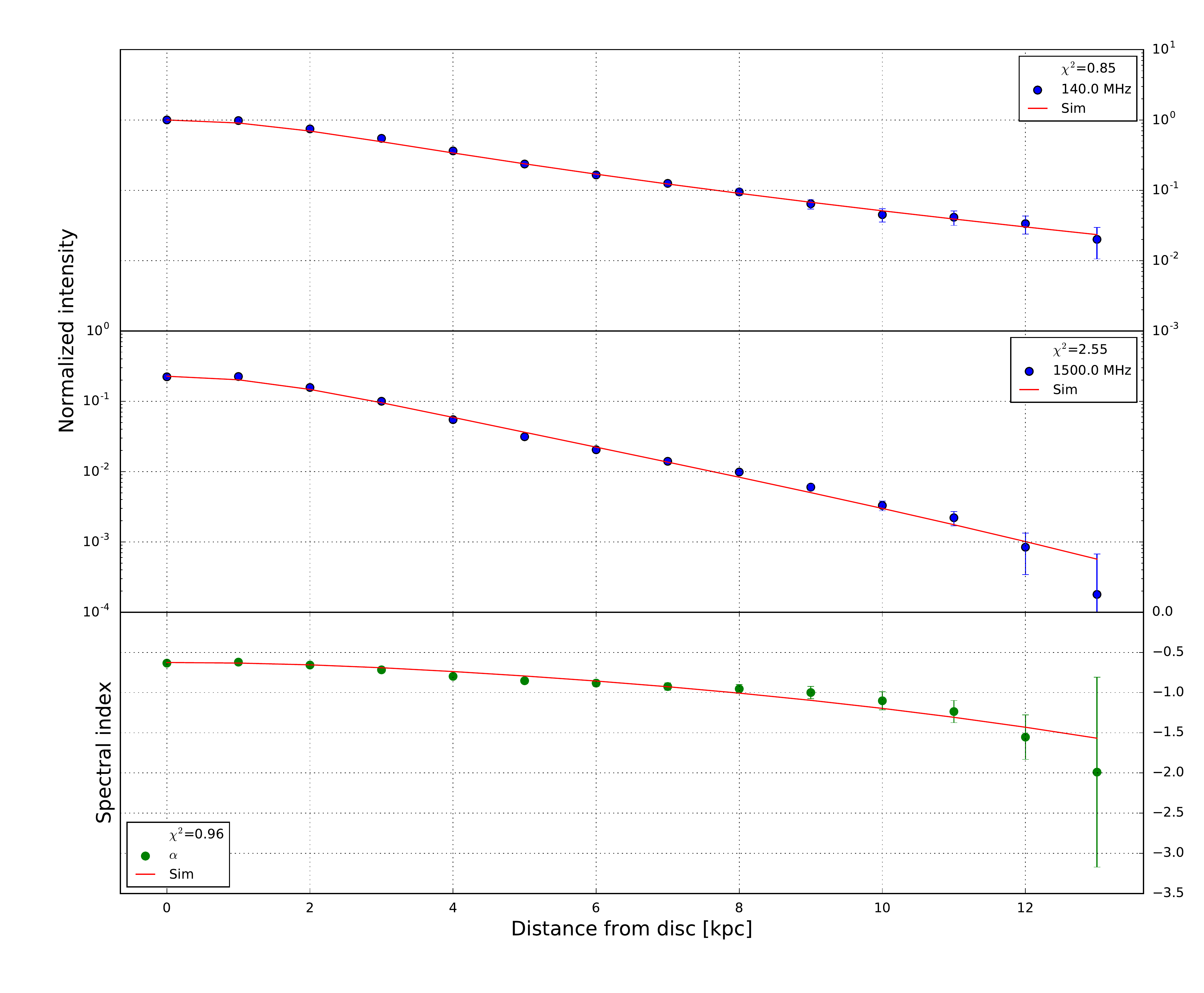}
\caption{Wind solution in the north-eastern quadrant of NGC~5775. From top to bottom, we show vertical profiles of the non-thermal intensities at $140~\rm MHz$ and 1500~MHz, respectively, and the non-thermal radio spectral index. Solid lines show the best-fitting advection model. The intensities were normalised with respect to the 140-MHz data point at $z=0~\rm kpc$.}
\label{figure:wind_ne}
\end{figure*}

\subsection{Application}\label{ss:application}

We have measured non-thermal intensities in vertical profiles as function of distance from the galactic midplane, as described in Section~\ref{subsec:vertdist}. In order to study the variation across the galaxy, we have divided the maps into four quadrants, and determined one intensity profile per quadrant. Values at large vertical distance from the plane were excluded at heights where the spectral index shows an apparent flattening (indicative of measurement errors or CR re-acceleration that would not be accommodated by our model), or measurement errors were excessively large. In practice we fitted the observed data up to $z=9\,\mathrm{kpc}$ in the SE and SW quadrants, and $z=13\,\mathrm{kpc}$ in the NE and NW quadrants. The profile for each quadrant was fitted with quasi-1D CR transport models of pure CR advection. For this, we implemented our wind model as described in Section~\ref{ss:wind_model} in the {\sc spinnaker} code \citep{heesen_etal_2016,heesen_etal_2018}. We then fitted the advection models to our data. We held the magnetic field strength in the midplane ($B_0$) constant at the value determined in Section~\ref{subsec:spix}. We fitted the flux tube scale height ($z_0$) and the flux tube opening parameter ($\beta$). We also fit the CR$e^-$ injection power-law index ($\gamma$) and the critical velocity ($v_{\rm c}$). These two parameters mostly fix the radio spectral index profile, with $\gamma$ related to the spectral index in the midplane and the critical velocity establishing the vertical profile of advection speed. 

\begin{table}
\caption{Best-fitting wind solutions.}
\label{table:wind}
\begin{tabular}{l cccc}
\hline
Parameter & NE & NW & SE & SW \\
\hline
$B_0$ (\si{\micro}G)$^a$ & $13.0$ & $13.0$ & $19.0$ & $19.0$ \\
$v_{\rm c}$ ($\rm km\,s^{-1}$)$^{b}$ & $310^{+70}_{-50}$ & $310^{+110}_{-60}$ & $380^{+80}_{-60}$ & $320^{+80}_{-60}$ \\
$z_{\rm c}$ (kpc)$^{c}$ & $0.7^{+0.8}_{-0.6}$ & $<0.1$ & $0.2^{+0.3}_{-0.1}$ & $<0.1$ \\ 
$z_0$ (kpc)$^d$ & $8.8^{+1.5}_{-1.2}$ & $7.0^{+1.5}_{-1.5}$ & $6.9^{+0.7}_{-0.7}$ & $4.2^{+0.6}_{-0.6}$  \\
$\beta$$^e$ & $1.8^{+0.5}_{-0.4}$ & $1.2^{+0.5}_{-0.2}$ & $1.8^{+0.2}_{-0.2}$ & $1.3^{+0.15}_{-0.1}$ \\
$\gamma$$^f$ & $2.20^{+0.1}_{-0.1}$ & $2.4^{+0.1}_{-0.1}$ & $2.1^{+0.08}_{-0.06}$ & $2.2^{+0.07}_{-0.06}$ \\
$\chi_\nu^2$$^g$ & $1.5$ & $1.9$ & $3.7$ & $2.0$ \\
\hline
\end{tabular}
\textbf{Notes.} \\
(a) Total magnetic field strength in the disk (fixed);\\
(b) Wind speed at the critical point; \\
(c) Vertical height of the critical point; \\
(d) Scale height of the flux tube (see Eq.~\ref{eq:flux_tube});  \\
(e) Power-law index for the flux tube (see Eq.~\ref{eq:flux_tube}); \\
(f) CRe injection spectral index; \\
(g) Reduced $\chi^2$. \\
Solutions are in the north-eastern (NE), north-western (NW), south-eastern (SE), and south-western (SW) quadrants.
\end{table}

\begin{figure*}
\centering
\includegraphics[width=\hsize]{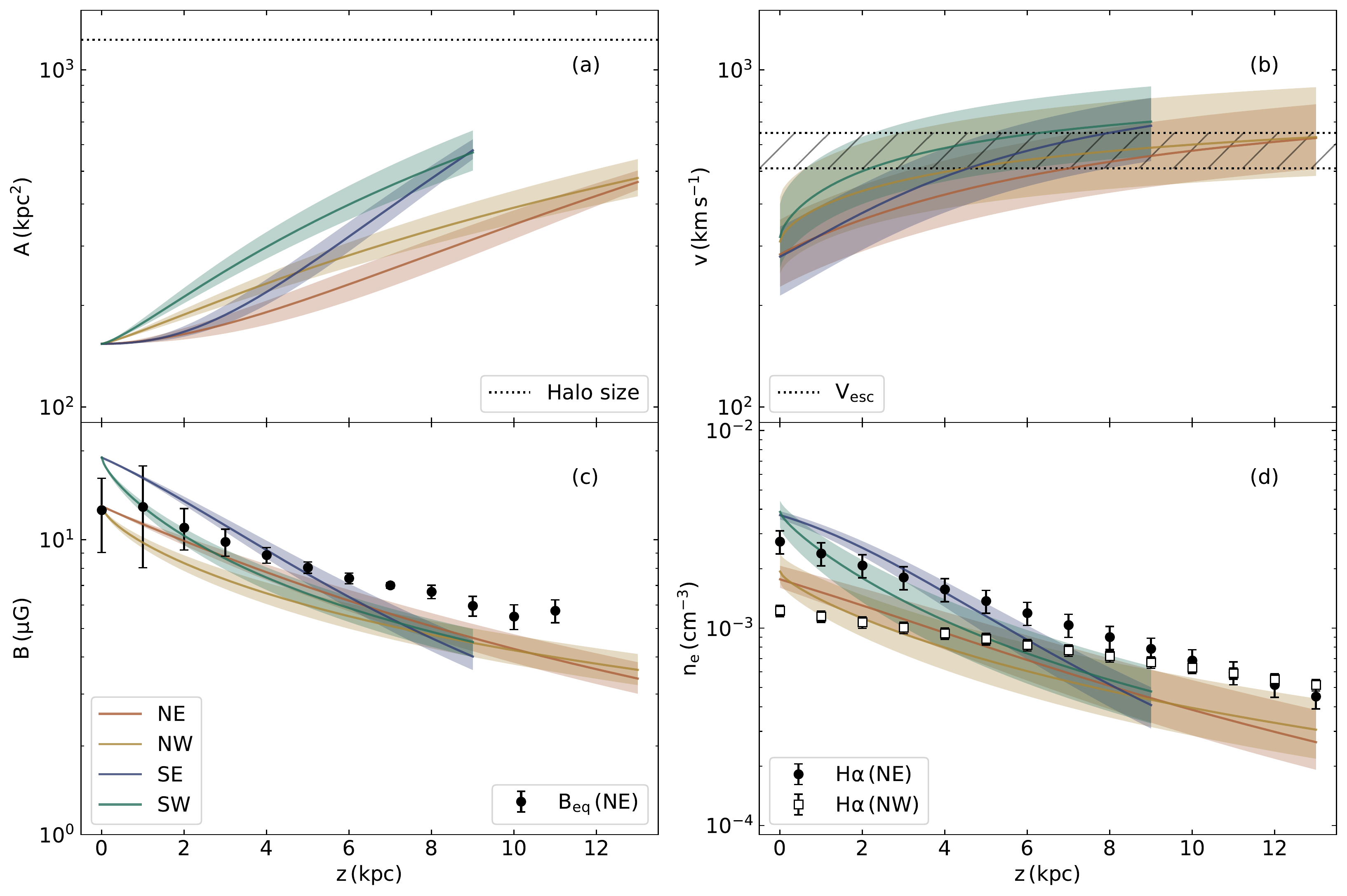}
\caption{Vertical variation of the physical parameters for the wind solution in each of the four quadrants of NGC~5775 separately. Solid lines show the best-fitting solutions while the shaded areas indicate 1$\sigma$ uncertainties. (a) Cross-sectional area of the outflow; the dotted horizontal line shows the cross sectional area corresponding to the maximum detected radial extent of the halo. (b) Advection velocity; the dotted lines bounding the hashed region indicate the lower and upper bounds for the expected range of escape velocities (as described in the text). (c) Magnetic field strength; filled black circles show the equipartition field strength in the NE quadrant. The other quadrants have similar values and are not shown for clarity. (d) Hot phase thermal electron density; data points show H$\alpha$ measurements by \citet{boettcher_etal_2019} scaled with a cloud volume filling factor of $f_{\rm cl}=0.05$ (filled black circles for the NE quadrant, and open black squares for the NW quadrant).}
\label{figure:n5775_val}
\end{figure*}

\subsection{Results}

\subsubsection{Advection velocity}
\label{sss:advection_velocity}

We find that our model fits the data reasonably well, with reduced chi-squared ($\chi_\nu^2$) values between $1.5$ and $3.7$. In Figure~\ref{figure:wind_ne}, we illustrate the best-fitting model in the NE quadrant; the best-fitting model for each of the other three quadrants is included in Appendix~\ref{a:wind_models}. Table~\ref{table:wind} presents the best-fitting parameters in all four quadrants. The flux tube model can account for the observed intensities, where we found that $\beta$ is between $1.2$ and $2.3$. This is in good agreement with $\beta=2.0$, so that the cross-sectional area generally increases as $\propto z^2$ (see also Section~\ref{sss:windgeometry}). We note that it may be difficult to reconcile the expanding geometry of this model with the boxy shape of the observed radio emission. Further modeling may be required to understand this apparent conflict. We discuss this issue further in Section~\ref{section:discussion}.

In Figure~\ref{figure:n5775_val}, we present the physical conditions in the outflow as implied by the fitted model. The area increases as can be expected for an expanding flow. We assumed an initial outflow radius of $r_0=7~\rm kpc$, approximately half the radius of the star-forming disc. The velocity starts at $200$--$400~\rm km\,s^{-1}$ at some height near the midplane, although we note that diffusion and streaming should dominate very close to the midplane; see Section~\ref{ss:comp_prev_models}. The velocity profile goes through the critical point within $1.0~\rm kpc$ distance from the disc. The wind then accelerates and reaches a velocity of $400$--$900~\rm km\,s^{-1}$ at the maximum vertical height from the disc that we considered for the model fits, namely 9~kpc (SE/SW) and 13~kpc (NE/NW) as described in Section~\ref{ss:application}. The wind velocity profile is nearly linear for $z\lesssim z_0$, and at larger heights the acceleration decreases. In previous work \citep{miskolczi_etal_2019,schmidt_etal_2019} we also found linear velocity profiles by empirically fitting different profile shapes. It is encouraging to see that our new wind model produces such profiles without any fine tuning (see Appendix~\ref{as:linearisation_of_the_wind_solution} for a more formal proof).

\subsubsection{Wind geometry}
\label{sss:windgeometry}

\begin{figure*}
    \includegraphics[width=\hsize]{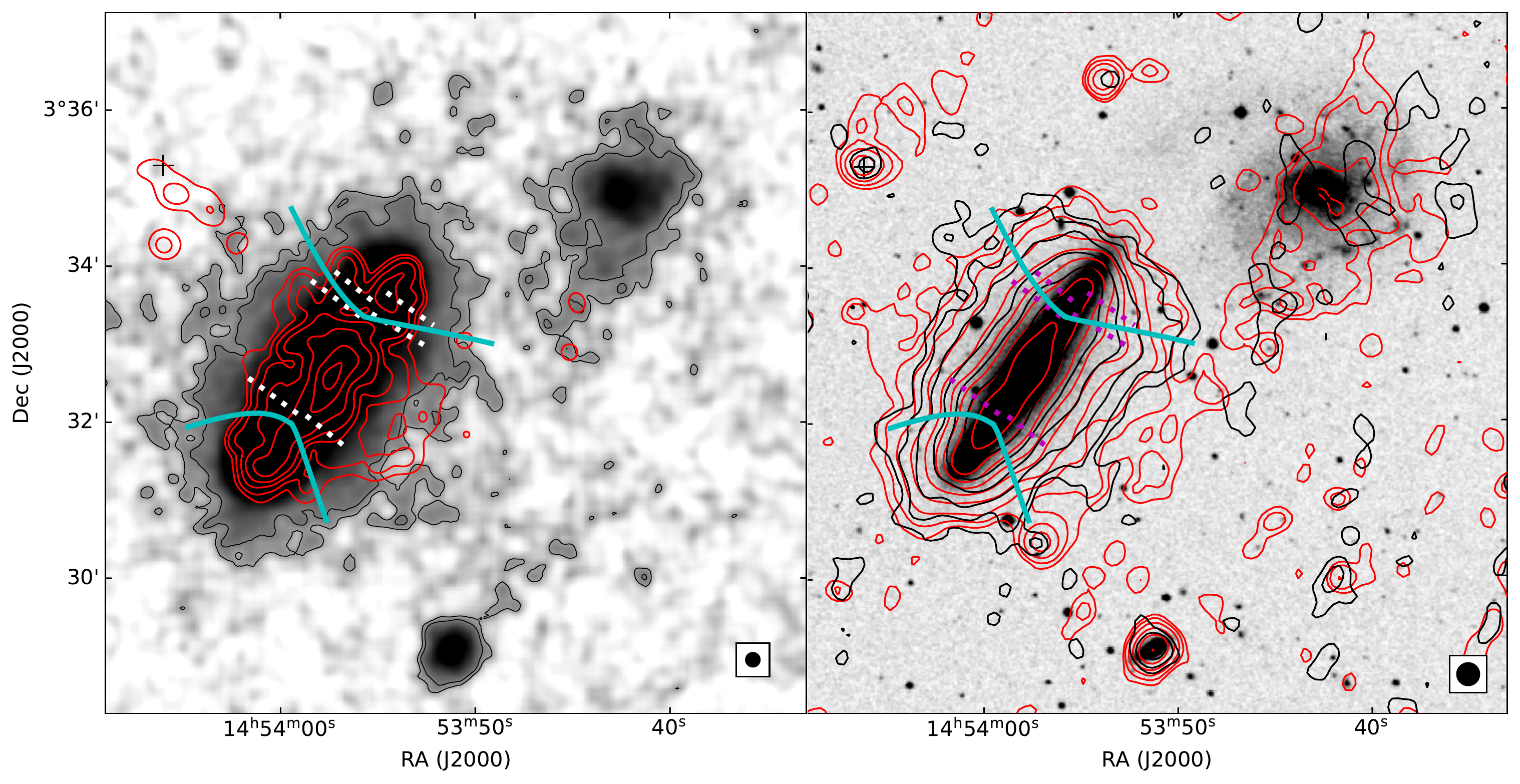}
    \caption{Flux tube model in the context of multi-wavelength images of NGC~5775. Left: {\it WISE} W4 (22~$\umu$m) image in greyscale presented with a log stretch, overlaid with black contours at 4 and $6\sigma$, and red contours from the adaptively-smoothed {\it Chandra} X-ray ($0.3$--$1.5$~keV) image from \citet{li_etal_2008}; the contours are at (2, 2.5, 3.5, 5, 8, 15, and 25)~$\times10^{-3}~\rm cts~s^{-1}~arcmin^{-2}$. NGC~5774 is not clearly detected in X-ray because that region was imaged with low sensitivity; see \citet{li_etal_2008}. Right: DSS2 $R$-band image of NGC~5775 and NGC~5774, overlaid with contours from the radio continuum images at 140~MHz (black) and 1500~MHz (red). For both radio images, the thermal contribution from NGC~5775 has not been subtracted; the contours start at 2$\sigma$ where $\sigma$ is the noise level reported in Table~\ref{table:images}, and increase by powers of two. In both panels, H$\alpha$ filament locations are plotted with dotted lines. The periphery of the fitted flux tube is indicated with cyan curves. The radio source near an X-ray `blob', discussed in the text, is marked with a plus. The galaxy to the south of NGC~5775 is IC~1070. The beam size of the {\it WISE} and radio continuum images is shown in the bottom right corner of each panel.}
    \label{fig:cone}
\end{figure*}

To illustrate the region of influence of the flux tube, we have overplotted the periphery of the wind model (as bounded by the run of $A(z)$; see Figure~\ref{figure:n5775_val}) on multi-wavelength images of NGC~5775 in Figure~\ref{fig:cone}. It is striking that the flux tube shape is approximately symmetric across the plane, although this was not constrained in the fitting process. The lateral expansion is clearly more rapid in the SE and SW quadrants, as captured by the $z_0$ and $\beta$ model parameters, and reflecting the typically smaller exponential scale height of the radio continuum emission on that side of the disc. Signs of a widescale wind are clearly visible in the infrared and X-ray images. The X-ray distribution displays a superbubble morphology in the southern halo \citep[see also][and specifically their Figure 9]{li_etal_2008}. The radial extent of the flux tube is reasonably consistent with the appearance of the infrared and X-ray distribution, although these were not used to constrain our model, and generally tends to enclose vertical features. It is possible that the prominent H$\alpha$ filaments mark the walls of the outflowing wind, and that they indicate entrainment of warm ionized material (see also Section~\ref{section:mass_loss_rate}).

We also note the appearance of the X-ray `blob' to the north-east of NGC~5775, originally identified by \citet{li_etal_2008}. Our radio images recover the possibly-associated radio source (marked with a plus in Figure~\ref{fig:cone}), with a spectral index $\alpha\approx-0.6$. This value is more suggestive of a distant radio galaxy rather than old plasma associated with the NGC~5775 wind, but we do not attempt to draw strong conclusions on the origin of this emission feature, nor do we comment further on this X-ray feature here. Also visible in Figure~\ref{fig:cone} is a clear indication of the radio continuum bridge connecting NGC~5775 and NGC~5774. As noted in Section~\ref{section:data}, this bridge is better reproduced in images generated in such a way as to emphasize larger angular scales than we have studied in this paper.

\subsubsection{Magnetic field strength}
\label{sss:magnetic_field_strength}

The magnetic field strength is determined by an interplay between the expanding area (increasing radius) and the increasing wind velocity. Our model field strengths can be well described with exponential functions having scale heights between 4 and 9~kpc. The CRs are in approximate energy equipartition everywhere. This can be understood in the following way. For the magnetic field strength we assumed $B\propto r^{-1}v^{-1}$, so that the magnetic energy density scales as $u_{\rm B}\propto r^{-2}v^{-2}$. The CR number density scales without adiabatic losses as $n\propto r^{-2}v^{-1}$, which is just the application of the continuity equation~\eqref{eq:continuity} for the CRs. With adiabatic losses included, the CR number density (per energy interval) scales as $n\propto (vA)^{({\gamma +2)/3}}$ \citep[e.g.,][]{baum_etal_1997}. Hence, the number density decreases slightly more.

We have compared our model field strengths with equipartition values derived in Section~\ref{subsec:spix}. We find that our field strengths are slightly (few \si{\micro}G) below the equipartition values. This is because our model describes the shape of the CR$e^-$ spectra in an improved fashion that admits the possibility of curved spectra, whereas the equipartition values assume a power-law spectrum. Even if the CR$e^-$ spectrum in the halo steepens due to energy loss, the spectrum of the total CRs will probably still be a power law, but with an uncertain slope, which increases the uncertainty of the equipartition estimate.

\subsubsection{Thermal electron density}
\label{sss:thermal_electron_density}

At the critical point, the advection speed is equivalent to the composite sound speed (equation~\ref{eq:composite_sound_speed}). Since we know the energy densities of the thermal hot gas and the CRs, we can calculate the corresponding pressures using $P_{\rm g}=(\gamma_{\rm g}-1)u_{\rm g}$ and $P_{\rm c}=(\gamma_{\rm c}-1)u_{\rm c}$, respectively. The energy density of the thermal hot halo gas is $u_{\rm g}=4\times 10^{-12}~\rm erg\,cm^{-3}$ \citep[][and Wang 2020, private communication]{li_etal_2008}, and for the CRs we assume energy equipartition with $u_{\rm c}=u_{\rm B}$, where the magnetic energy density in the galactic midplane is $u_{\rm B}=B_0^2/(8\upi)=(7$--$14)\times 10^{-12}~\rm erg\,cm^{-3}$. With $P_{\rm g}=3\times 10^{-12}~\rm dyn\,cm^{-2}$ and  $P_{\rm c}=(2$--$4)\times 10^{-12}~\rm dyn\,cm^{-2}$ the thermal hot gas and the CRs are approximately in pressure equilibrium. With equation~\eqref{eq:composite_sound_speed} we can then calculate the gas density $\rho$ at the critical point and then we employ the continuity equation~\eqref{eq:continuity} to calculate the density elsewhere. The thermal electron density in the hot phase is then $n_{\rm e}=\rho/(2\bar \umu m_{\rm u})$ with a mean molecular weight of $\bar \umu = 0.65$ and $m_{\rm u} = 1.67\times 10^{-24}~\rm g$. The resulting vertical thermal electron density profiles are shown in Fig.~\ref{figure:n5775_val}.

The thermal electron density falls off with height and can be well described by an exponential function with scale heights between $3.0$ and $6.5$~kpc. The electron density starts with a volume density of $n_{\rm e}=(2$--$4)\times 10^{-3}~\rm cm^{-3}$ near the midplane and falls off to approximately $(0.2$--$0.4)\times 10^{-3}~\rm cm^{-3}$ at the edge of the halo. These values are indicative of the hot ionized medium (HIM). It would be interesting to draw a connection to the warm ionized medium (WIM) which is not expected to play an important role in driving the wind, but is likely entrained in the flow. However, it is hard to make a prediction for the WIM thermal electron density due to uncertainties in the pressure balance and filling factors of each phase. Nevertheless, we note that \citet{boettcher_etal_2019} found for the thicker (`halo') of two vertical components midplane electron densities of $\sqrt{f_{\rm cl}}n_{\rm e}=0.05~\rm cm^{-3}$ with a scale height of $3.6\pm 0.2~\mathrm{kpc}$ in the south-west of NGC~5775 and $\sqrt{f_{\rm cl}}n_{\rm e}=0.02~\rm cm^{-3}$ with scale height of $7.5\pm 0.4$~kpc in the north-east. Here, $f_{\rm cl}$ is the WIM cloud volume filling factor. These electron densities were measured at a range of vertical distances along a slit perpendicular to the disc, and we have compared the measurements with our models as presented in Fig.~\ref{figure:n5775_val}. The scale heights are in good agreement, and the average electron density values are similar to the model prediction for the HIM if we adopt a low volume filling factor (around $f_{\rm cl}=0.05$), which is a plausible comparison for typical values of the relative gas phase pressures \citep[e.g.,][]{ferriere_2001}.

\section{Discussion}
\label{section:discussion}

\subsection{Wind-driven mass loss rate}
\label{section:mass_loss_rate}

Since the continuity equation is fulfilled, we can make an estimate of the mass loss rate that the galaxy experiences by calculating the mass flux at the critical point. The advection speed may increase to high enough values that the gas will escape from the galaxy. The escape velocity does not depend strongly on the properties of the dark matter halo, with $v_{\rm esc}=(2.6$--$3.3)v_{\rm rot}$ \citep{veilleux_etal_2005} for a truncated isothermal sphere halo model. In this picture and with the observed $v_{\rm rot}$ listed in Table~\ref{table:n5775properties}, the escape velocity for NGC~5775 is between 510 and $650~\rm km\,s^{-1}$, and so the escape velocity is likely exceeded already at the detected edge of the halo. Since the wind would accelerate even further \citep[e.g.,][]{breitschwerdt_etal_1991}, our model predicts that the CR-driven wind largely meets the escape condition, as indicated in Figure~\ref{figure:n5775_val}. The total mass-loss rate (both above and below the plane, combined) is $\dot M = (3.2$--$7.6)\epsilon\times 10^{26}~\rm g\,s^{-1}$, which equates to $\dot M=(5$--$12)\epsilon~\rm M_{\sun}\,yr^{-1}$, where the parameter $\epsilon$ was introduced in Section \ref{subsec:Physical_picture} to indicate the efficiency of entraining gas into the wind. While we may not expect all of the hot gas to participate in the wind, on the other hand we do expect some warm ionized and neutral clouds to be entrained. These factors would need to be included in the overall value of $\epsilon$.

\begin{figure}
\centering
\includegraphics[width=\hsize]{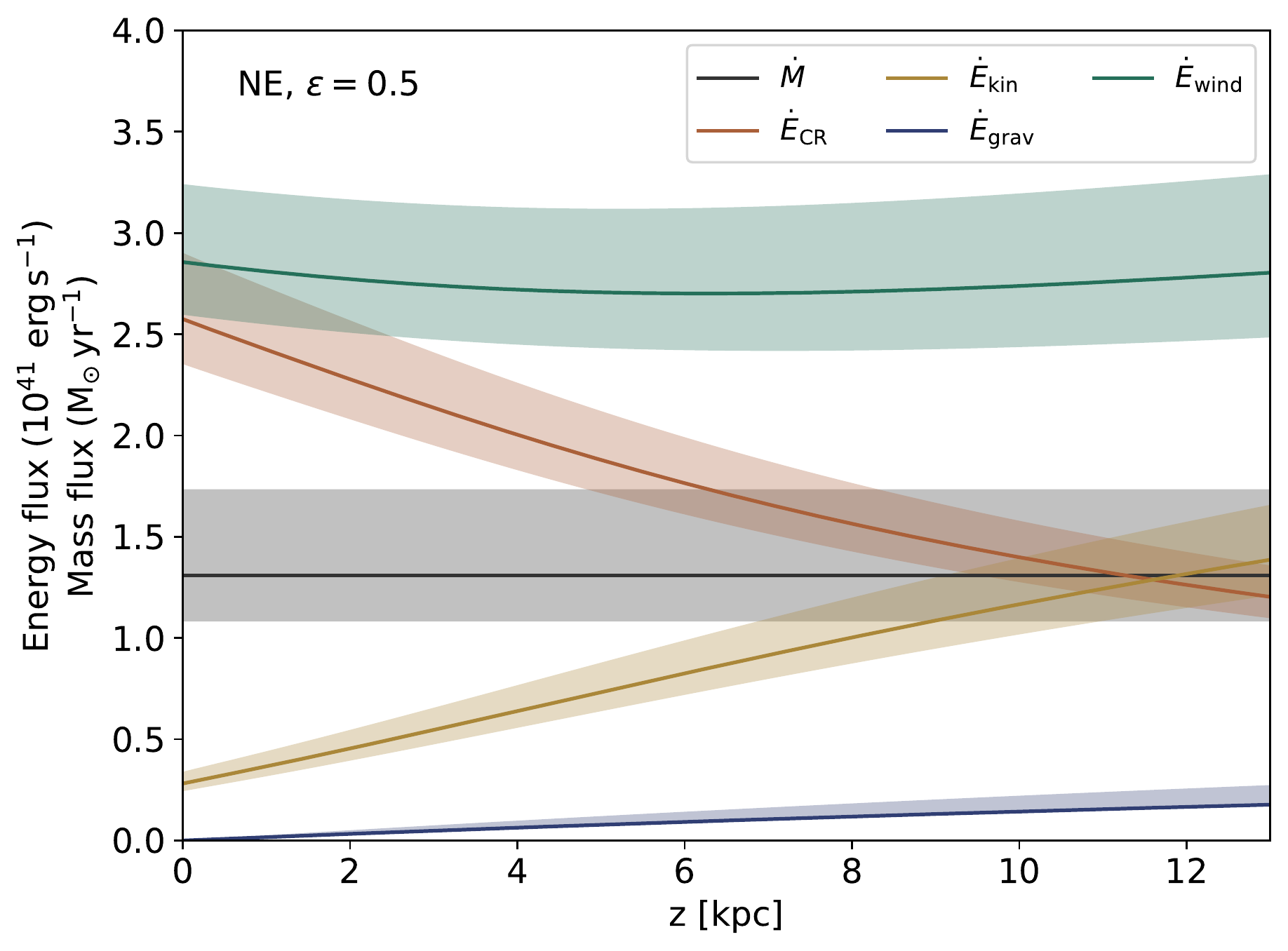}
\caption{Energy and mass fluxes in the north-eastern halo (extrapolated to an entire hemisphere) for an entrainment efficiency of $\epsilon = 0.5$. The mass flux $\dot M$ is constant due to the continuity equation~\eqref{eq:continuity}. The CR energy flux ($\dot E_{\rm CR}$) decreases, the energy of which is transferred into the kinetic energy flux ($\dot E_{\rm kin}$) of the gas and into the work lifting the gas in the gravitational potential ($\dot E_{\rm grav}$). The total energy flux of the wind $\dot E_{\rm wind}$ is approximately constant.}
\label{figure:flux_ne}
\end{figure}

The derived mass loss-rate for $\epsilon=1$ should be broadly interpreted as an upper limit. We have explored whether it is energetically feasible for a star formation-driven wind to support such a large mass flux. The adopted SFR from H$\alpha$+IR is $7.56~\mathrm{M_{\sun}\,yr^{-1}}$ (see Table~\ref{table:n5775properties}), so that the core-collapse supernova rate is $\nu_{\rm SN} \simeq 0.09~\rm yr^{-1}$ \citep{murphy_etal_2011}. Assuming that each supernova injects $10^{51}~\rm erg$ of kinetic energy into the interstellar medium, of which 10 per cent is converted into CR acceleration \citep[e.g.,][and references therein]{rieger_etal_2013}, the energy injection rate for CRs is $\dot{E}_\mathrm{CR}=3\times 10^{41}\,\mathrm{erg\,s^{-1}}$. This is slightly lower than our measured CR energy fluxes of $(6$--$12)\times 10^{41}\,\mathrm{erg\,s^{-1}}$. However, the alternative SFR derived from the $1.4$-GHz radio continuum--SFR relation is as high as $22.9~\mathrm{M_{\sun}\,yr^{-1}}$ using the total radio continuum flux measured from our images and the SFR calibration from \citep{murphy_etal_2011}, which would provide sufficient energy. The wind would hence lead to a steady state where the produced CRs are transported in the wind and energy losses within the galaxy are small.

So that the wind is energetically feasible for a given value of the entrainment efficiency $\epsilon$, the energy flux of the wind,
\begin{equation}
\dot E_{\rm wind} = \dot E_{\rm CR} + \dot E_{\rm kin} + \dot E_{\rm grav},
\end{equation}
should be approximately constant as function of height above the disc. The kinetic energy flux of the wind can be calculated as $\dot{E}_\mathrm{kin} = \frac{1}{2} \dot{M} v^2$. At the edge of the halo, we find $\dot{E}_\mathrm{kin} = (3$--$10) (\epsilon/0.5) \times 10^{41}\,\mathrm{erg\,s^{-1}}$. The energy flux that is needed in order to lift the gas in the gravitational potential is $\dot E_{\rm grav} =\dot M \int g {\rm d}z$, which is only about 10 per cent of the kinetic energy flux. In Fig.~\ref{figure:flux_ne}, we show the mass and energy fluxes as an example for the north-eastern halo (the remaining quadrants are shown in Appendix~\ref{ass:remaining_quadrants}). We find that the energy flux is indeed approximately constant for $\epsilon\approx0.5$, which is the value that we adopt henceforth. This is the global value with some spatial variation possible. Because our wind model makes some simplified assumptions, especially the constant composite sound speed, further uncertainties will be introduced in the exact profile shape of the vertical energy fluxes. In particular, the cosmic-ray energy density is expected to decrease faster with height due to adiabatic cooling. While this is in part compensated by the entrainment efficiency, the actual profile shape will be different for a self-consistent model. Nonetheless, in this way we estimate the mass-loss rate to be $\dot{M} \approx (3$--$6) (\epsilon/0.5)\,\mathrm{M_{\sun}\,yr^{-1}}$. The mass-loading efficiency is then $\eta\equiv \dot{M}/SFR$, and we find $\eta = (0.4$--$0.8)\,(\epsilon/0.5)$ where we have again used our adopted SFR estimate.  Mass-loading efficiencies of order unity are expected for stellar feedback-driven winds which include CRs \citep{samui_et_al_2010,mao_ostriker_2018}.

\subsection{Observational diagnostics}
\label{ss:observational_diagnostics}

\subsubsection{Lagging haloes}
\label{sss:lagging_haloes}

\begin{figure}
\centering
\includegraphics[width=\hsize]{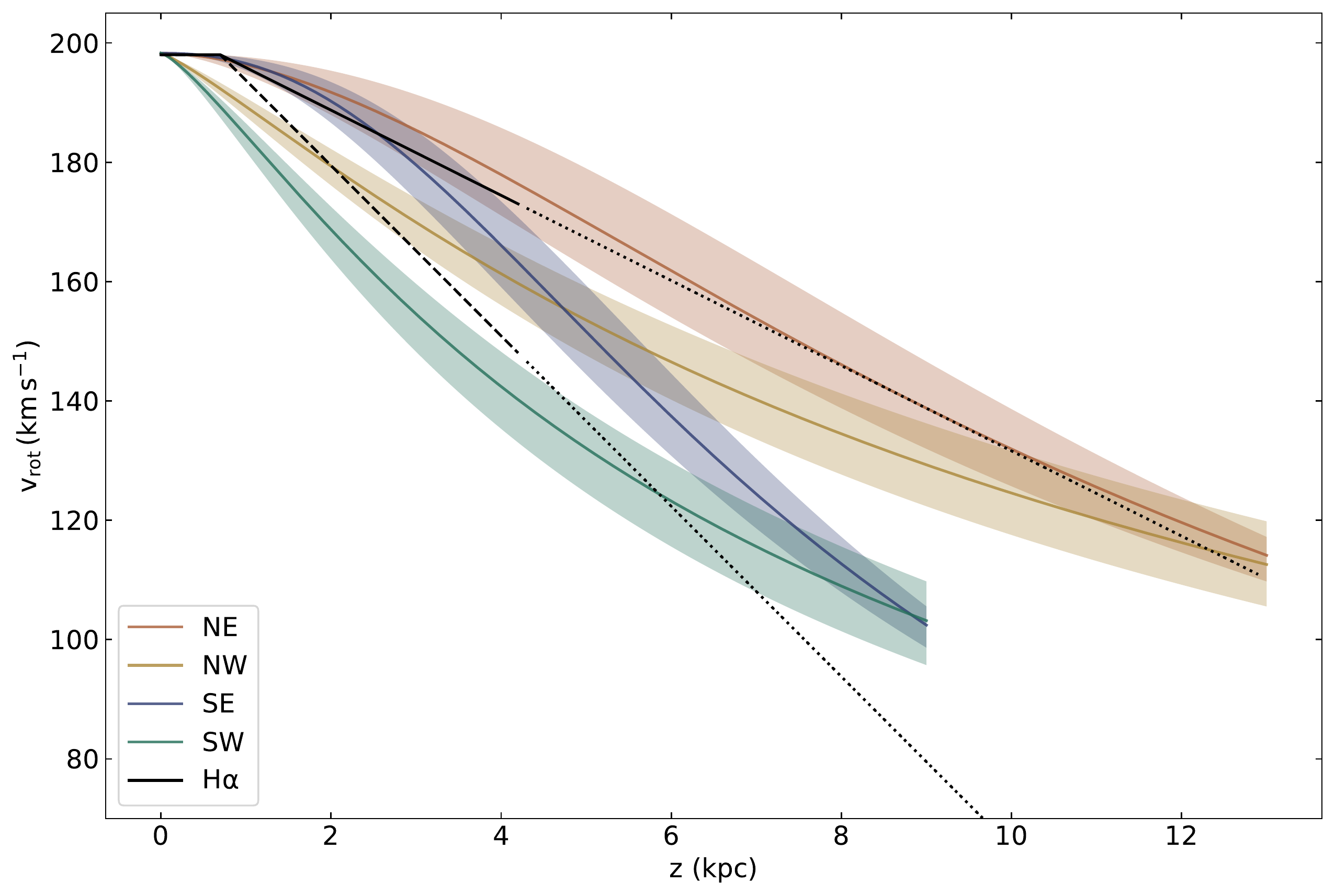}
\caption{Predicted rotation speed as a function of distance from the disc at radii between $7$~kpc (at $z=0~\rm kpc$) and $12-16~\rm kpc$ (at the maximum distance; see Fig.~\ref{figure:n5775_val}). The solid black line represents the best-fitting kinematic model determined on the basis of Fabry--Perot interferometry of the H$\alpha$ emission line by \citet{heald_etal_2006}, over the range of $z$ considered in that study. The dashed black line indicates the steeper gradient in rotation velocity that \citet{heald_etal_2006} indicated is appropriate in localized regions. Dotted black lines extrapolate both determinations of the lag to larger vertical distance.}
\label{figure:n5775_rot}
\end{figure}

Our one-dimensional model does not allow us to explicitly incorporate rotation, but we note that an interesting expected consequence of the expanding outflow model is that the rotation speed will decrease with increasing height above the galactic midplane due to conservation of angular momentum. Under that assumption
\begin{equation}
    l = v_\mathrm{rot} r,    
\end{equation}
and so the rotation speed $v_{\rm rot}$ decreases in inverse proportion to the outflow radius. The predicted rotation speed as a function of distance from the midplane is shown in Fig.~\ref{figure:n5775_rot} in each of the quadrants. As in other calculations we use the midplane flux tube radius $r_0=7~\mathrm{kpc}$, this time for the launching radius, but the outflow radius increases in the halo. The rotation speed decreases with a typical gradient between $-12$ and $-6\,\mathrm{km\,s^{-1}\,kpc^{-1}}$. These values are in very good agreement with H$\alpha$ measurements by \citet{heald_etal_2006} who found a vertical gradient in rotation velocity of $-7~\rm km\,s^{-1}\,kpc^{-1}$ (corrected for the distance assumed in this paper), with evidence for steeper gradients in localized areas. More recent measurements by \citet{boettcher_etal_2019} found steeper gradients of up to $-25~\rm km\,s^{-1}\,kpc^{-1}$ along a single vertical slice, cautioning, however, that they may overestimate the true rotational gradient. The question is whether the warm ionized gas is really participating in the outflow. Nevertheless, the observed velocity gradient is at least in good agreement with our model.

Previous kinematic measurements of a sample of galaxies including NGC~5775 initially indicated an apparent correlation between the scale height of thick disc gas and the corresponding vertical lag in rotational velocity \citep[][for a small sample of three galaxies only]{heald_etal_2007}. This tentative correlation was later disputed by \citet{zschaechner_etal_2015} on the basis of their larger sample. In the flux tube model used here, the scale height of the thermal gas is linked through the lateral expansion of the wind to a vertical decrease in rotational velocity, and thus we expect some sort of relationship between them. 

We can derive the expected value of the rotational gradient for our model from the linearisation of the radius around $z=z_0$ (Appendix~\ref{ass:rotation_velocity_gradient}):
\begin{equation}
    \diff{}{z}v_{\rm rot} = -\frac{v_{\rm rot}}{2\sqrt{2}z_0},
\end{equation}
where $v_{\rm rot}$ is the maximum rotation speed in the disc midplane. Hence, we expect velocity gradients between $-18$ and $-8~\rm km\,s^{-1}\,kpc^{-1}$ similar to values already derived. Additionally, we can approximate the thermal electron density scale height as $h_{\rm e}\approx z_0/[2\log(2)]$ (Appendix~\ref{ass:thermal_electron_density_scale_height}). Taken together, this means that the rotational gradient can be expressed in terms of the extraplanar diffuse ionized gas (eDIG) scale height as:
\begin{equation}
     \diff{}{z}v_{\rm rot} \approx -\frac{v_{\rm rot}}{4\sqrt{2}\log(2)h_{\rm e}}.
\end{equation}
Hence, the expected gradient would be $51~\mathrm{km\,s^{-1}}$ per unit eDIG scale height for a rotation curve with $v_\mathrm{rot}=200\,\mathrm{km\,s^{-1}}$. This is somewhat higher than the typical values of $20~\mathrm{km\,s^{-1}}$ per scale height as had been suggested by \citet{heald_etal_2007}, but provide a closer match to the observations than the predictions from ballistic models \citep{collins_rand_2002,fraternali_binney_2006} which typically underestimate the rotational gradient by a factor of a few \citep[e.g.,][]{heald_etal_2007}. 

\subsubsection{X-shaped halo magnetic fields}
\label{sss:x_shaped_halo_magnetic fields}

Another important diagnostic is radio continuum polarization. NGC~5775 is the prototype of galaxies showing an `X-shaped' halo magnetic field, where the magnetic field orientation on kpc-scales shows a characteristic shape, reminiscent of an `X'. These magnetic field structures are possibly connected to the outflow, and indeed an outflowing wind is a crucial element in efficiently amplifying an `X'-shaped field through the dynamo mechanism \citep[see, e.g.,][and references therein]{woodfinden_etal_2019}. This association is what we have assumed in our model. The cone half-opening angle is:
\begin{equation}
    \theta_{1/2} = \tan^{-1} \left (\frac{r-r_0}{z}\right ).
\end{equation}
For $\beta = 2$ the cone half-opening angle is nearly constant, otherwise it is a function of distance from the disc. We find cone half-opening angles of $\theta_{1/2}\approx 25\degr$, resulting in cone-opening angles of $50\degr$. This is a fairly large opening angle, although \citet{soida_etal_2011} find an even larger opening angle of $80\degr$. Hence the model prediction seems to be at least comparable to observations, and indeed we would not expect perfect agreement since the impact of the wind on the field depends on the velocity profile of the flow, the field strength and its orientation with respect to the flow. Also, the observations may be affected by superposition of the disc magnetic field onto the projected field orientation. A proper modelling of the magnetic field structure is, however, beyond the scope of this paper.

On the other hand, our modelling does seem to be in possible conflict with the boxy shape of the radio halo, especially at LOFAR frequencies (Section~\ref{subsec:vertdist}). Our adopted flux tube geometry would seemingly naturally imply a more bi-conical shape at large vertical distances from the midplane. As seen in Figure~\ref{fig:halpha}, that is clearly not the radio morphology of NGC~5775. It is possible that the boxiness is accentuated through the interaction with NGC~5774, particularly at the northern edge where the radio continuum morphology is particularly flattened. We have used the flux tube model as a straightforward way to consistently treat the vertical variation of key quantities, in the framework of a 1D model. The geometry is clearly an approximation of a disc-scale outflow, and one way to reconcile the model with the boxy halo morphology could be to picture a more widespread wind with average vertical speed increasing outward. It may ultimately become necessary to adopt a two-dimensional model framework, or even a system that incorporates differential rotation. Although beyond the scope of the present paper, future work could attempt to resolve the apparent morphological conflict by selecting a different model configuration that accommodates similar behaviours of physical parameters to Figure~\ref{figure:n5775_val} but with an eye toward matching the apparent cylindrical geometry of the synchrotron emitting region above the midplane. We leave this development to future work.

\subsection{Comparison with previous models}
\label{ss:comp_prev_models}

Compared with our earlier models \citep[as used by e.g.,][]{mulcahy_etal_2018}, our new wind model features the following improvements in describing the propagation of CRs: (i) the vertical decrease of the magnetic field strength can now be described coherently through an accelerated, laterally expanding wind; (ii) there is no longer any need for an ad-hoc exponential magnetic field distribution; and (iii) the increase of the advection speed is the result of the conservation of total energy. Hence, with our new model, we have reduced the free parameters in the fit to the flux scale height $z_0$ and the speed at the critical point $v_\mathrm{c}$. In previous work \citep[e.g.,][]{schmidt_etal_2019,miskolczi_etal_2019}, we fitted the exponential magnetic field scale height and the velocity scale height separately. There was some freedom in selecting these parameters, and so we chose the CRs to be in approximate energy equipartition with the magnetic field. This is now also a result of the model without requiring any subsequent fine adjustments. The work presented by \citet{mulcahy_etal_2018} and \citet{schmidt_etal_2019} should now be revisited with our new model.

We nevertheless explored whether our previous models can also adequately fit the present data. We tested a constant wind speed and found an equivalently good match to the data. However, as already found in our previous work, the resulting magnetic field strength would be significantly below the equipartition value. The models presented by \citet{heesen_etal_2018} for 12 galaxies (including NGC~5775) result in a median departure from equipartition by about an order of magnitude at a characteristic height of $10$~kpc, and the constant-velocity fit results using the new data in the present paper yield a similar problem. Such a large discrepancy is incompatible with observations of radio halos \citep[e.g.,][]{duric_1990,mora_etal_2019,seta_beck_2019}. It is possible to obtain approximate energy equipartition even with a constant wind in the flux tube geometry; however, the resulting pressure gradient would have to be exactly balanced by the gravitational acceleration, so that such an arrangement is inherently unstable. For these reasons, we therefore favour models incorporating an increasing advection speed. We found that an exponential magnetic field together with a linearly increasing advection speed is also sufficient to fit the data equally well, but requires one additional free parameter in form of the velocity scale height, and subsequent tuning as described above.  This shows again, as was already pointed out in \citet{miskolczi_etal_2019}, that with LOFAR we are now are able probe the halo out to distances in excess of 10 kpc from the midplane and yet the vertical spectral index profiles have shallow slopes and are almost linear. Such behaviour can be well modelled with CR advection as opposed to (pure) diffusion \citep[for such a case, see][]{heesen_etal_2019b}, indicating the presence of winds.

Nevertheless, there remain a few limitations, which we discuss next. Our pure advection model can not explain why a galactic wind develops in the first place. For this, CR diffusion and/or streaming are highly important \citep{wiener_17a,farber_18a}. Our models assume that once a galactic wind is present, advection will dominate over diffusion everywhere except quite close to the disc for typical diffusion coefficients of $\sim 10^{28}~\rm cm^2\,s^{-1}$. The case of streaming may be important as well, but the magnetic field structure and radio spectral index distribution suggest that this could be a very local process (compare with Fig.~\ref{fig:spix}). Our 1D model applied to averaged data in entire quadrants does not take these relatively small-scale effects into account in an explicit way. Nevertheless, we acknowledge that the unknown contribution from CR (anisotropic) diffusion and streaming  can potentially significantly impact our results since we only take CR advection into account. Clearly, future modelling should revisit these issues and attempt an even more physically motivated description. For instance, diffusion and streaming depend on the local magnetic field structure, which can be modelled on large scales ($\gtrsim1$~kpc) using the CHANG-ES polarization data \citep{krause_etal_2020}; on smaller scales, constraining aspects such as the degree of anisotropy in the turbulent component of the magnetic field will require new deep, high angular resolution observations such as with the Square Kilometre Array \citep[SKA; see for example][]{beck_etal_2015}.

\section{Conclusions}\label{section:conclusions}

In this paper, we have presented new LOFAR 140~MHz radio continuum observations of the nearby starburst galaxy NGC~5775. The resulting map is the most sensitive of this galaxy to date below 1~GHz. We supplemented this low-frequency image with a new multi-configuration 1.5~GHz map from the VLA CHANG-ES survey. In preparing these images we took particular care to match angular resolution and sensitivity that was optimised for the detection and analysis of vertically extended radio continuum emission. On the basis of these images, we subtracted a nominal contribution from thermal emission using combined H$\alpha$ and \emph{Spitzer} 24-\si{\micro}m data. Following this correction, the pair of images was used to calculate non-thermal radio spectral indices over a decade in frequency span, as well as estimates of the equipartition magnetic field strength. We then determined vertical profiles of the non-thermal intensity and fitted them to determine exponential scale heights.

The resulting vertical intensity and non-thermal spectral index profiles were fitted with a newly expanded CR transport model, building on our previous 1D advection models by implementing a simple wind model. We assumed a tunable flux tube geometry with an approximately hyperboloidal form, together with an assumption of iso-thermal flow, and an exponentially decreasing vertical component of the gravitational acceleration. These assumptions were adopted in order to coherently develop a physically-motivated vertical variation of key quantities relevant to the propagation and energy loss of the CRs, while introducing few free parameters. Our new CR transport model can explain many of the observed features in the radio halo. On the other hand, we point out that our simplified wind model is not yet self-consistent and incorporates an unphysical source of energy which leads to an overestimate of the wind acceleration in the halo. An alternative, fully self-consistent wind model is introduced in Appendix~\ref{as:cosmic_ray_driven_wind}, which qualitatively matches the model presented here. The alternative model will be explored in more detail and directly confronted with observational data in a forthcoming paper.

These are our main conclusions:
\begin{itemize}
    \item The morphology of the radio halo of NGC~5775 exhibits a boxy appearance, especially at the lowest frequencies that are probed by LOFAR. We detect radio emission up to about 13~kpc from the midplane at both 140 and 1500~MHz. 
    \item Thermal emission contributes a relatively large proportion of the continuum radiation observed from NGC~5775, due to the high level of widespread star formation. At locations with prominent ongoing star formation activity, up to 57 per cent (23 per cent) of the continuum radiation is thermal in nature at 1500 (140)~MHz, and with typical values elsewhere of 12 per cent (4 per cent).
    \item The non-thermal continuum emission is well characterised by a single exponential distribution, with typical thick disc scale heights of $2$--$3\,\mathrm{kpc}$ depending on frequency and location in the galaxy. The scale height at 140~MHz is larger than or equal to that at 1500~MHz, with an average scale height ratio of $1.2\pm0.3$.
    \item The non-thermal spectral index distribution steepens away from the midplane as observed in previous work. Our new images show that `channels' of shallower spectral index coincide with prominent H$\alpha$ filaments, with corresponding aligned extensions of the ordered magnetic field. This may be indicative of localised regions of ongoing CR streaming.
    \item We find that the average equipartition magnetic field strength across NGC~5775 is \mg{8}. The field is typically 50 per cent higher in the midplane of the southern side of the disc than in the north. Exponential scale heights of the magnetic field distribution are highly varied, but on average are $\approx18\,\mathrm{kpc}$. 
    \item Our new wind model does a good job of matching the vertical distribution of non-thermal radio continuum emission and corresponding steepening of the non-thermal spectral index. The model accounts for varying magnetic field strength, thermal gas density, and wind speed with height, but without requiring explicit ad-hoc vertical variation of each parameter separately.
    \item The modeled scale height of the thermal hot gas is 5~kpc, similar to measurements of the WIM scale height from the H$\alpha$ emission line. The midplane thermal electron density is $(2$--$4)\times 10^{-3}~\mathrm{cm^{-3}}$, in approximate agreement with X-ray observations.
    \item The expanding flow results in a decreasing magnetic field strength with approximately the same scale height as suggested by the data, but with typical values somewhat below the equipartition strengths derived directly from the observations. This suggests that the latter may be overestimated due to the different spectral behaviour of CRe and total CRs.
    \item Due to angular momentum conservation, the lateral expansion of the wind has the consequence that the rotational velocity will decrease with height. We find a predicted vertical gradient in rotational velocity with values between $-(6$--$12)~\mathrm{km\,s^{-1}\,kpc^{-1}}$. This is in excellent agreement with observations by \citet{heald_etal_2006}, who found an average gradient from H$\alpha$ spectroscopy of $-7~\mathrm{km\,s^{-1}\,kpc^{-1}}$ (corrected for the distance assumed in this paper), and noted indications of higher values in localized regions.
    \item With our new wind model, we are able to calculate the mass-loading of the outflow. For both hemispheres taken together, we find a mass-loss rate of $\dot{M} = (\epsilon/0.5)(3$--$6)~\mathrm{M_{\sun}\,yr^{-1}}$. Hence, the mass-loading efficiency is $\eta\approx (0.4$--$0.8\,(\epsilon/0.5)$, where $\epsilon$ is the efficiency of entraining gas in the wind.
    \item We find a predicted opening angle of the flow around $\approx 50\degr$. If the magnetic field lines follow the flow of the plasma, we should be able to compare this with polarization measurements; however, \citet{soida_etal_2011} found a somewhat larger opening angle of $\approx 80\degr$.
    \item A possible shortcoming of our model is that it would tend to suggest a biconical appearance for the radio continuum halo of NGC~5775, whereas we observe a boxy morphology. Future models should consider how to reconcile the vertical variation of the key parameters in the context of a vertical flow confined to an overall cylindrical structure.
\end{itemize}

Our results highlight the unique opportunities that are now enabled by low-frequency radio continuum observations in the study of galactic winds. Together with ongoing enhancements in the modeling of CR propagation, we are developing a clearer picture of the impact of winds on the structure and kinematics of star forming galaxies. NGC~5775, as a well-studied case with one of the most extended known radio continuum haloes, provides the opportunity to test and develop new aspects of our models, so that we can apply them in the future to galaxies with lower star-formation rates.

The prospects are particularly promising as we enter the era of SKA precursor surveys and draw closer to the construction of the SKA itself. Extending this study to observations with higher sensitivity and angular resolution will soon be possible for galaxies observable from the Southern hemisphere through the use of the Australian SKA Pathfinder (Hotan et al., submitted), MeerKAT \citep{jonas_etal_2016}, and ultimately over a very broad range of frequency with the SKA. The current proliferation of enhanced observational capability is set to unlock community ambitions to better understand magnetic fields and their role in the structure and evolution of galaxies \citep[e.g.,][]{heald_etal_2020}.

\section*{Data availability}

The data underlying this article will be shared on reasonable request to the corresponding author. The datasets were derived from sources in the public domain: 
\begin{itemize}
    \item The LOFAR data are available from the Long Term Archive (LTA; \url{https://lta.lofar.eu/}) under project code LC1\_046.
    \item The VLA data are available from the NRAO Archive Access Tool (AAT; \url{https://archive.nrao.edu/}) under project code 10C-119. Standard images released as part of CHANG-ES Data Release 1 \citep{wiegert_etal_2015} are available from the survey webpage (\url{https://www.queensu.ca/changes/}).
\end{itemize}

\section*{Acknowledgments}

This paper is based (in part) on data obtained with the International LOFAR Telescope (ILT) under project code LC1\_046. LOFAR \citep{vanhaarlem_etal_2013} is the Low Frequency Array designed and constructed by ASTRON. It has observing, data processing, and data storage facilities in several countries, that are owned by various parties (each with their own funding sources), and that are collectively operated by the ILT foundation under a joint scientific policy. The ILT resources have benefitted from the following recent major funding sources: CNRS-INSU, Observatoire de Paris and Universit\'e d'Orl\'eans, France; BMBF, MIWF-NRW, MPG, Germany; Science Foundation Ireland (SFI), Department of Business, Enterprise and Innovation (DBEI), Ireland; NWO, The Netherlands; The Science and Technology Facilities Council, UK; Ministry of Science and Higher Education, Poland.

The National Radio Astronomy Observatory is a facility of the National Science Foundation operated under cooperative agreement by Associated Universities, Inc.

G.~H. thanks Phil Edwards for useful feedback after a critical reading of the manuscript, and Tim Galvin for advice and help with tricky plotting issues. We would like to thank B.~P.~Brian Yu, for kindly providing us with their computer code of the cosmic ray-driven wind model. We thank the anonymous referee for a comprehensive review and for comments that led to substantial improvements to the paper.
MB acknowledges support from the Deutsche Forschungsgemeinschaft under Germany's Excellence Strategy - EXC 2121 "Quantum Universe" - 390833306.
This research was undertaken as an activity organized by the LOFAR Magnetism Key Science Project (MKSP; \url{https://lofar-mksp.org/}).

This research made use of Photutils, an Astropy package for detection and photometry of astronomical sources \citep{photutils}.

\bibliographystyle{mnras}
\bibliography{n5775}

\appendix

\section{Iso-thermal wind models}
\label{a:wind_models}

In this appendix, we provide further detail regarding the justification for our selected vertical gravitational acceleration (Section~\ref{ass:gravitational_potential}) and the derivation and solution of the wind model (Section~\ref{ass:derivation_of_wind_equation} and \ref{ass:solving_the_wind_equation}, respectively). Section~\ref{ass:useful_analytical_expressions} presents a few useful analytical expressions that come out of the solution of the wind equation. We also present the best-fitting wind models in the remaining three quadrants (Section~\ref{ass:remaining_quadrants}) to supplement the result for the NE quadrant as was shown in the main paper (Figure~\ref{figure:wind_ne}).

\subsection{Gravitational potential}
\label{ass:gravitational_potential}
\begin{figure}
\centering
\includegraphics[width=\hsize]{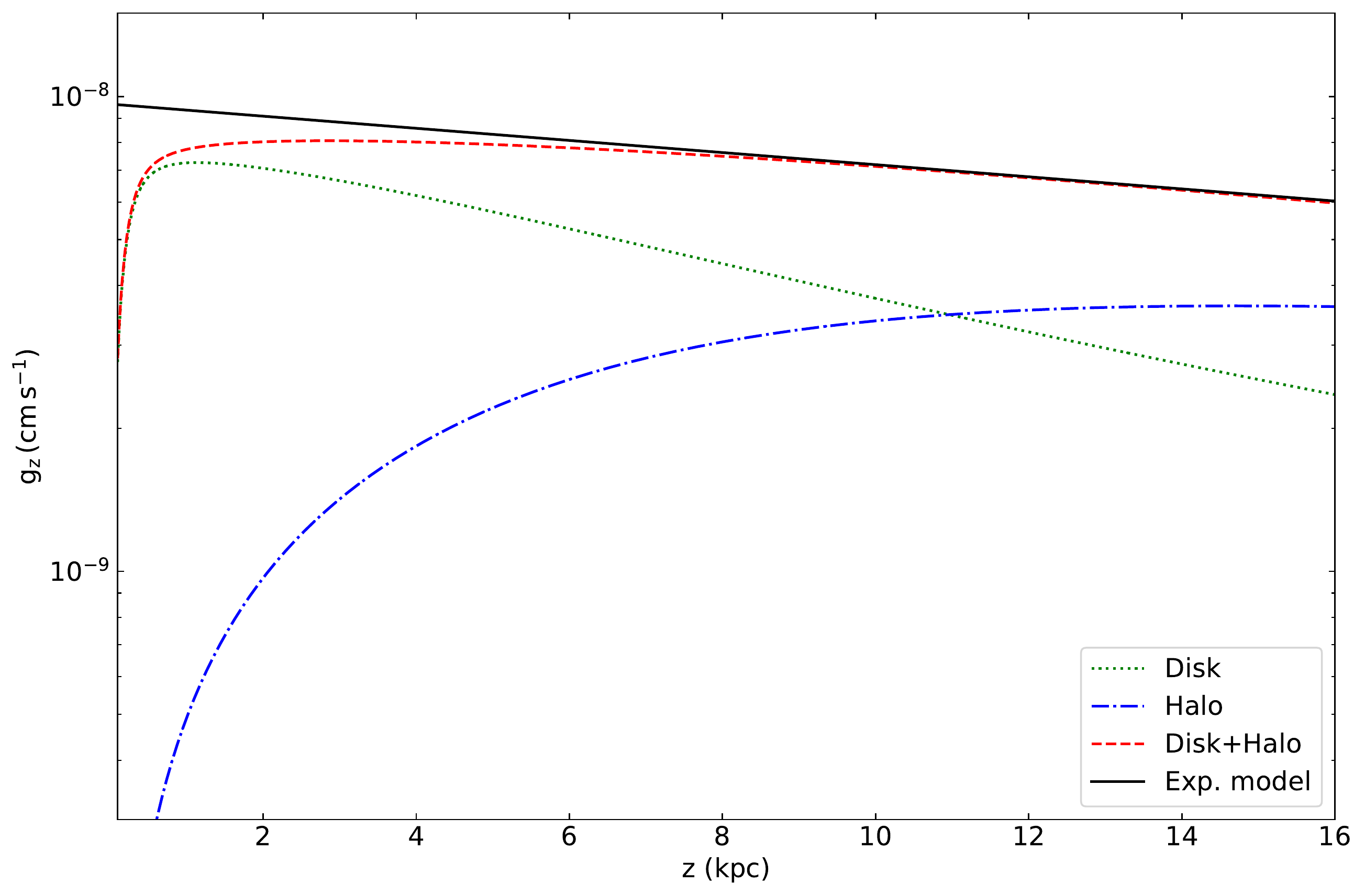}
\caption{Gravitational vertical acceleration in the Milky Way at a galactocentric radius of $R=8.5~\rm kpc$ as function of distance from the galactic midplane. The green dotted line shows the contribution from the disc, the blue dot-dashed line the contribution from the halo, and the red dashed line the sum of both. Equations adopted from \citet{wolfire_etal_1995}. The black solid line is our exponential model.}
\label{figure:n5775_grav}
\end{figure}
The vertical gravitational acceleration can be approximated by an exponential distribution. In this appendix, we motivate this approach. The gravitational potential is the superposition of the disc, bulge, and dark matter halo potentials. Since we do not have such detailed information for NGC~5775, we use a scaled version of the Milky Way gravitational potential as a template. We use the potentials as presented by \citet{wolfire_etal_1995}. The gravitational potential of the bulge is neglected since NGC~5775 has no noticeable bulge. Our approximation is:
\begin{equation}
    g = \frac{v_{\rm rot}^2}{2 r_0} \exp\left (\frac{-z}{h_{\rm grav}}\right ),
    \label{eq:grav}
\end{equation}
where we assume $v_{\rm rot}=225~\rm km\,s^{-1}$. 
Hence, the gravitational acceleration decreases exponentially with a scale height of $h_{\rm grav}=4r_0$. In Figure~\ref{figure:n5775_grav}, we show the vertical profiles of the vertical gravitational acceleration for the Milky Way. The disc potential dominates near the disc and the halo potential takes over as the dominant contribution at 10~kpc distance from the Galactic midplane. The sum of both contributions can be fitted reasonably well with our model from $z\approx 2~\rm kpc$ onwards. We tested that this holds also for only half of the solar radius, $R=4.25~\rm kpc$. For larger radii, this does not hold well anymore as the dark matter halo potential starts to dominate, which does not decrease as rapidly. 

In NGC~5775, we assume that the wind is launched within a galactocentric radius of $R=7~\rm kpc$. Hence, we scale the model in Equation~(\ref{eq:grav}) to our radius and rotation speed. 

\begin{figure*}
\centering
\includegraphics[width=\hsize]{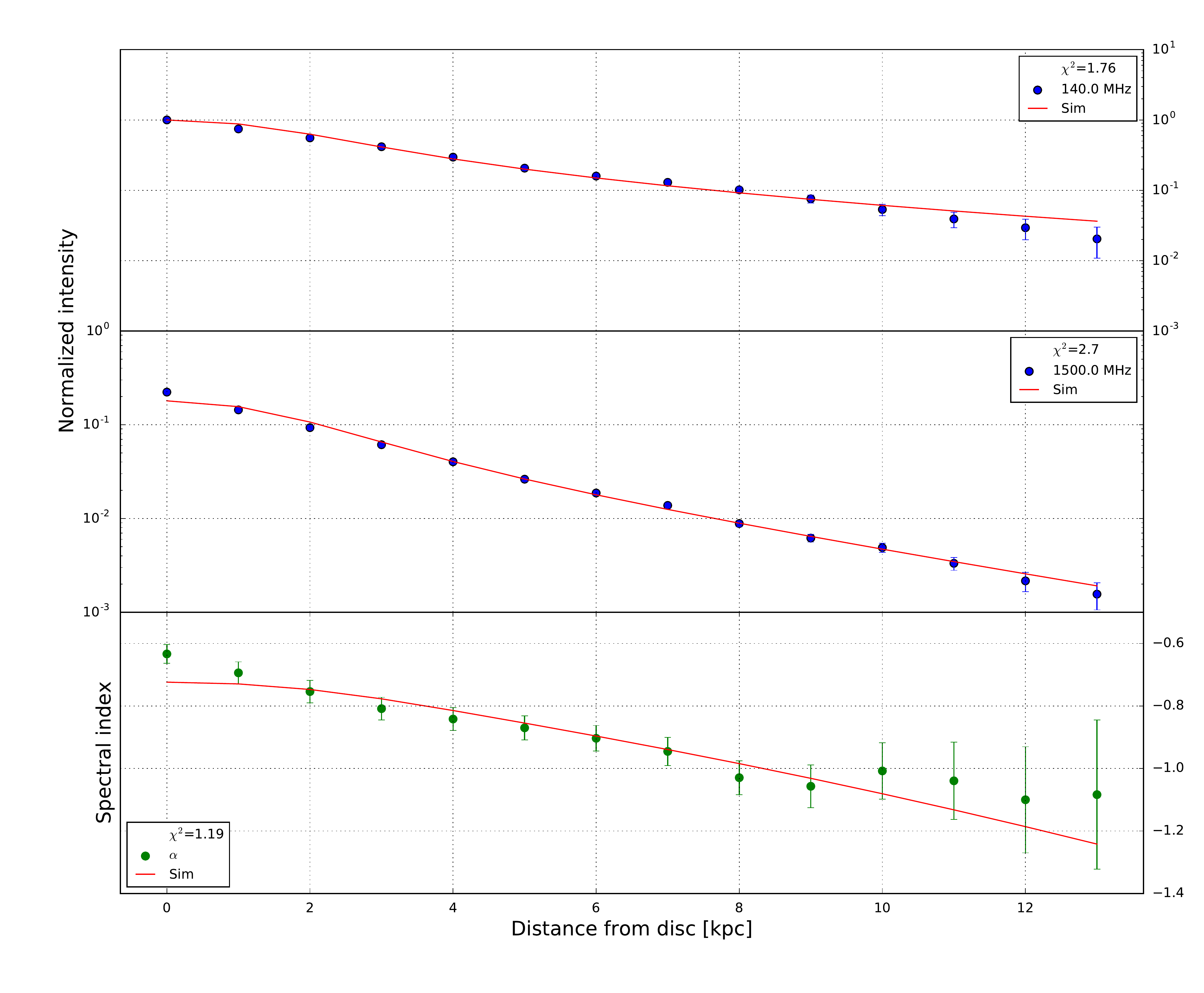}
\caption{Wind solution in the north-western quadrant of NGC~5775. From top to bottom, we show vertical profiles of the non-thermal intensities at $140~\rm MHz$ and $1.5~\rm GHz$, respectively, and the non-thermal radio spectral index. Solid lines show the best-fitting advection model. The intensities were normalised with respect to the 140-MHz data point at $z=0~\rm kpc$.}
\label{figure:wind_nw}
\end{figure*}

\begin{figure*}
\centering
\includegraphics[width=\hsize]{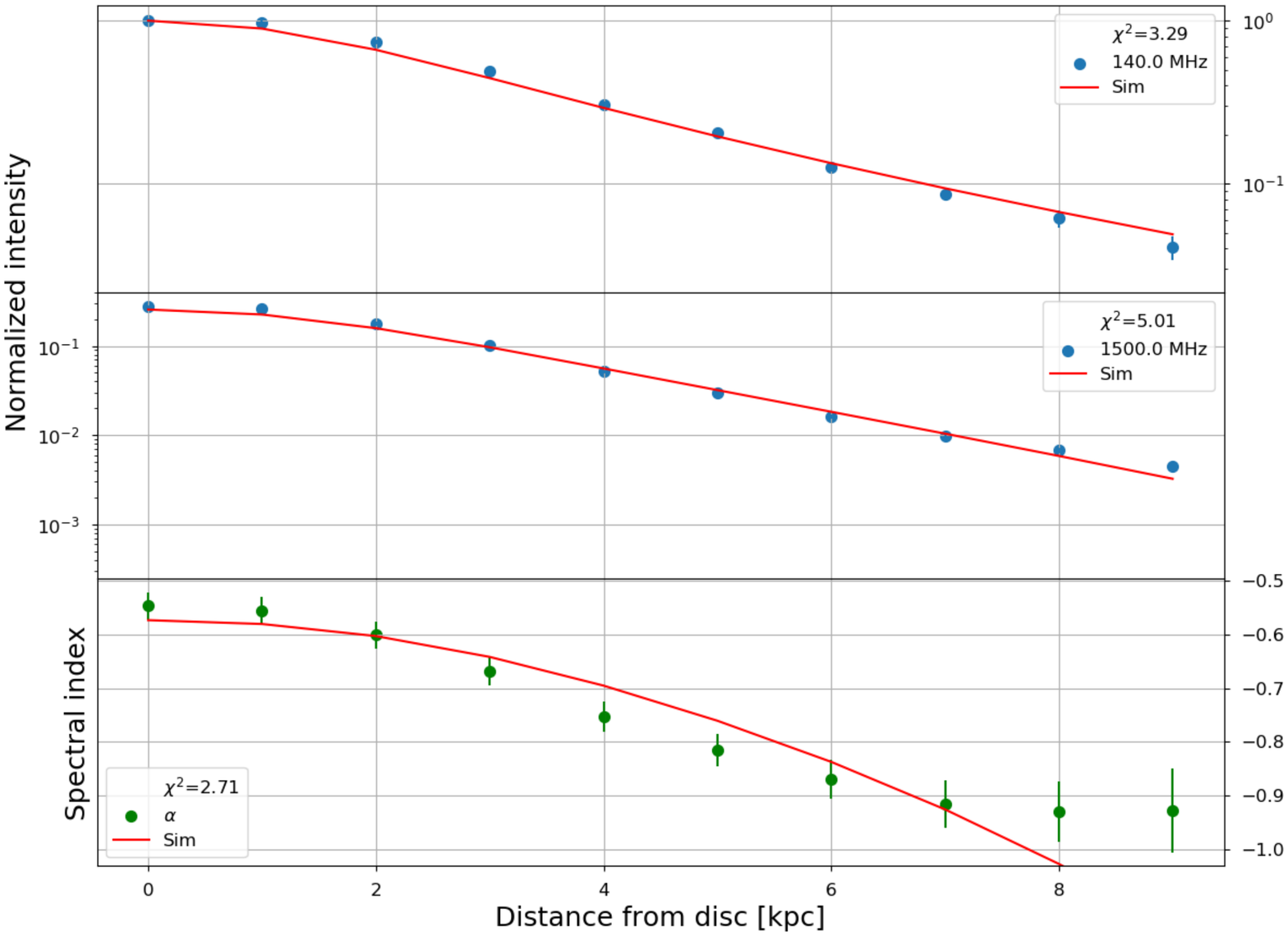}
\caption{Wind solution in the south-eastern quadrant of NGC~5775. From top to bottom, we show vertical profiles of the non-thermal intensities at $140~\rm MHz$ and $1500~\rm MHz$, respectively, and the non-thermal radio spectral index. Solid lines show the best-fitting advection model. The intensities were normalised with respect to the 140-MHz data point at $z=0~\rm kpc$.}
\label{figure:wind_se}
\end{figure*}

\subsection{Derivation of the wind equation}
\label{ass:derivation_of_wind_equation}
In this appendix, we derive the wind equation in a similar way as for the Parker-type winds \citep[see e.g.][for a derivation of this particular type of wind equation]{boyd_2003}. The only differences to the solar wind are that we assume the wind to be a stellar feedback-driven one with pressure contributions both from the thermal and CR gas (although that does not change the equation), the outflow geometry is not spherical but a flux tube geometry, and the gravitational acceleration falls off exponentially rather than with the square of the radius. 

The Euler (conservation of momentum) equation is:
\begin{equation}
  \rho v \frac{{\rm d}v}{{\rm d}z} = \frac{{\rm d}P}{{\rm d}z} - g\rho.
  \label{aeq:euler_equation}
\end{equation}
We introduce the critical velocity $v_{\rm c}^2=P/\rho$, so that:
\begin{equation}
    \diff{P}{z} = v_{\rm c}^2 \diff{\rho}{z}. 
\end{equation}
Plugging this into the Euler equation~\eqref{aeq:euler_equation}, we obtain:
\begin{equation}
    \rho v \diff{v}{z} = -v_{\rm c}^2 \diff{\rho}{z} - g\rho.
    \label{aeq:euler_rho}
\end{equation}
Plugging the expanding flux tube approximation,
\begin{equation}
    A = A_0 \left [1 + \left ( \frac{z}{z_0}\right )^{\beta}\right ],
    \label{aeq:flux_tube}
\end{equation}{}
into the continuity equation,
\begin{equation}
    \diff{}{z}\left ( \rho A v\right ) = 0,
    \label{aeq:continuity}
\end{equation}
we obtain the vertical density gradient:
\begin{equation}
    \diff{\rho}{z} = - \left [ 1 + \left (\frac{z}{z_0}\right )^{\beta} \right]^{-1} \frac{\beta z^{\beta-1}}{z_0^{\beta}} \rho - \frac{\rho}{v}\diff{v}{z}.
\end{equation}
Inserting the density gradient into equation~\eqref{aeq:euler_rho}, the Euler equation~\eqref{aeq:euler_equation} becomes:
\begin{equation}
    \rho v \diff{v}{z} = v_{\rm c}^2\left [ 1 + \left ( \frac{z}{z_0}\right )^{\beta} \right ]^{-1} \frac{\beta z^{\beta-1}}{z_0^{\beta}} \rho + \frac{v_{\rm c}^2\rho}{v} \diff{v}{z} - g\rho.
    \label{eq:euler_rho_expanded}
\end{equation}
The expanded Euler equation~\eqref{eq:euler_rho_expanded},  expressed in $\rho$, needs to be integrated. The flux tube approximation in equation~\eqref{aeq:flux_tube} can be integrated:
\begin{equation}
    \int \frac{z^{\beta-1}}{1+\left ( \frac{z}{z_0}\right )^{\beta}} {\rm d}z = \frac{z_0^{\beta}}{\beta}\log \left [1+\left (\frac{z}{z_0}\right )^\beta \right ].
\end{equation}
Hence, integration of the expanded Euler equation~\eqref{eq:euler_rho_expanded} leads to:
\begin{equation}
    \int  \left ( v-\frac{v_{\rm c}^2}{v} \right )\diff{v}{z} {\rm d}z = \int \left \{ \frac{\beta v_{\rm c}^2 z^{\beta-1}}{z_0^{\beta}} \left [ 1 + \left (\frac{z}{z_0}\right )^{\beta}\right ]^{-1} -g\right \} dz,
\end{equation}
which can be written as:
\begin{equation}
    \frac{1}{2}v^2 -v_{\rm c}^2\log(v) = v_{\rm c}^2 \log\left[1+\left (\frac{z}{z_0}\right )^{\beta} \right ]    - \int g {\rm d}z + C^{\prime},
\end{equation}
where $C^{\prime}$ is an integration constant. Changing to a suitable integration constant $C$, we find:
\begin{equation}
    \left (\frac{v}{v_{\rm c}}\right)^2 -\log \left(\frac{v}{v_{\rm c}}\right)^2 = 2\log \left[1+\left (\frac{z}{z_0}\right )^{\beta} \right ] - \frac{2}{v_{\rm c}^2}\int g {\rm d}z +C.
\end{equation}
Now we integrate the gravitational potential:
\begin{equation}
    \int g {\rm d}z = \int g_0 \exp\left (\frac{-z}{h_{\rm grav}} \right ) {\rm d}z = -g_0 h_{\rm grav}\exp\left( \frac{-z}{h_{\rm grav}}\right ).
\end{equation}
Thus, we finally obtain the `wind equation':
\begin{eqnarray}
    \left ( \frac{v}{v_{\rm c}}\right )^2 - \log\left ( \frac{v}{v_{\rm c}} \right )^2 & = & 2 \log\left [ 1+\left (\frac{z}{z_0}\right )^\beta \right ] \nonumber \\ & + & \frac{v_{\rm rot}^2h_{\rm grav}}{v_{\rm c}^2r_0}\exp\left ( -\frac{z}{h_{\rm grav}} \right )\nonumber \\ & + & C.
    \label{aeq:wind_equation}
\end{eqnarray}

The equation becomes undefined at the critical point where $v=v_{\rm c}$ and $z=z_{\rm c}$, so that the right-hand-side ($rhs$) of equation~\eqref{aeq:wind_equation} has to fulfil $rhs=1$. This means the integration constant is:
\begin{eqnarray}
    C & = & 1 - 2\log\left [ 1+\left (\frac{z_{\rm c}}{z_0}\right )^\beta\right ] \\
  & - &   \frac{v_{\rm rot}^2h_{\rm grav}}{v_{\rm c}^2r_0}\exp\left ( -\frac{z_{\rm c}}{h_{\rm grav}} \right ).
  \label{aeq:integration_constant}
\end{eqnarray}
So that equation~\eqref{aeq:wind_equation} is defined everywhere, $rhs\geq 1$ needs to be fulfilled. Hence, the right-hand-side has a minimum of $rhs=1$ at the critical point $z=z_{\rm c}$.

\subsection{Solving the wind equation}
\label{ass:solving_the_wind_equation}
In order to solve the wind equation~\eqref{aeq:wind_equation}, we first have to determine the location of the critical point. The derivative of $rhs$ is:
\begin{equation}
    \diff{}{z}(rhs) = \frac{2\beta z^{\beta-1}}{1+\left(\frac{z}{z_0}\right)^{\beta}}z_0^{-\beta} - \frac{v_{\rm rot}^2}{v_{\rm c}^2r_0}\exp\left (\frac{-z}{h_{\rm grav}}\right ).
\end{equation}
This becomes:
\begin{equation}
    \diff{}{z}(rhs)= 2\beta z^{\beta-1} \frac{v_{\rm c}^2r_0}{v_{\rm rot}^2}\exp\left ( \frac{-z}{h_{\rm grav}} \right ) - (z_0^{\beta} + z^{\beta}).
\end{equation}{}
Then, for the minimum at the critical point $z=z_{\rm c}$ we require:
\begin{equation}
    \diff{}{z}(rhs) = 0,
\end{equation}
which becomes:
\begin{equation}
    z_{\rm c}^{\beta} - 2\beta z_{\rm c}^{\beta-1} \frac{v_{\rm c}^2 r_0}{v_{\rm rot}^2}\exp\left ( \frac{z_{\rm c}}{h_{\rm grav}}\right ) + z_0^{\beta}=0.
\end{equation}
This can be solved iteratively with the Newton method, where:
\begin{equation}
    f_{\rm crit} = z_{\rm c}^{\beta} - 2\beta z_{\rm c}^{\beta-1} \frac{v_{\rm c}^2 r_0}{v_{\rm rot}^2}\exp\left ( \frac{z_{\rm c}}{h_{\rm grav}}\right ) + z_0^{\beta},
\end{equation}{}
and
\begin{eqnarray}
    f_{\rm crit}^{\prime} & = &\beta z_{\rm c}^{\beta-1} -2\beta \left (\frac{v_{\rm c}}{v_{\rm rot}}\right )^2 r_0 \exp\left ( \frac{z_{\rm c}}{h_{\rm grav}}\right ) \nonumber \\
    & & \left [(\beta-1)z_{\rm c}^{\beta-2} +  \frac{v_{\rm c}^{\beta-1}}{h_{\rm grav}} \right ].
\end{eqnarray}
Then,
\begin{equation}
    z_{\rm c}^{i+1} = z_{\rm c}^i - \frac{f_{\rm crit} }{f_{\rm crit}^{\prime}}.
\end{equation}{}
The wind equation is solved in an equivalent way, first defining a dimensionless wind speed:
\begin{equation}
    \tilde{v} \equiv \frac{v}{v_{\rm c}}.
\end{equation}{}
Then the wind equation becomes:
\begin{equation}
    \tilde{v}^2 - 2\log(\tilde{v}) - rhs =0.    
\end{equation}
Defining:
\begin{equation}
    f_{\rm wind} =\tilde{v}^2 - 2\log(\tilde{v}) -rhs,
\end{equation}
and:
\begin{equation}
    f_{\rm wind}^{\prime} = 2\tilde{v} - \frac{2}{\tilde{v}},
\end{equation}
the solution can be again found iteratively:
\begin{equation}
    \tilde{v}^{i+1} = \tilde{v}^{i} - \frac{f_{\rm wind}}{f_{\rm wind}^{\prime}}.
\end{equation}{}
The solution converges quickly, which then gives the wind velocity.

\begin{figure*}
\centering
\includegraphics[width=\hsize]{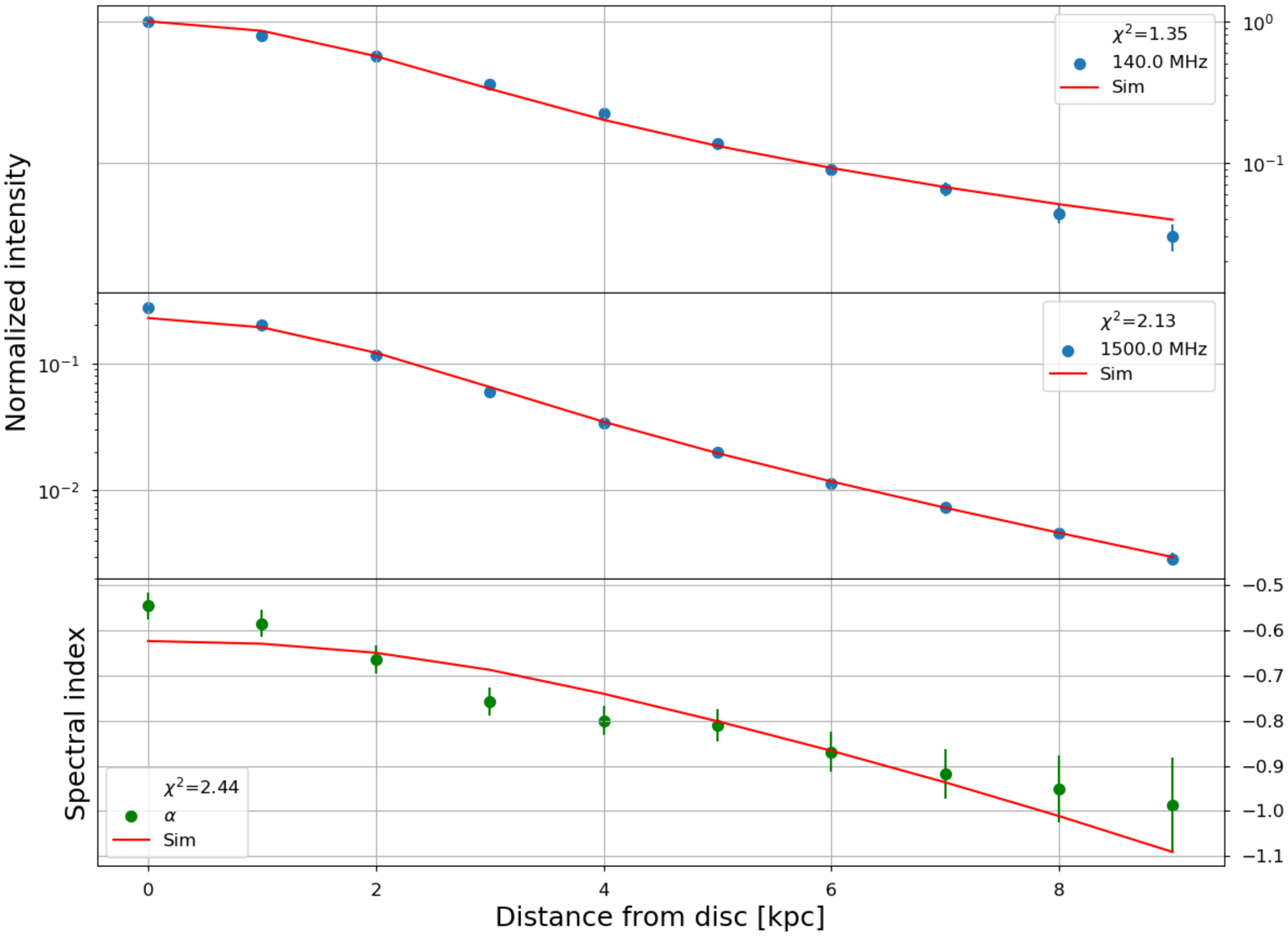}
\caption{Wind solution in the south-western quadrant of NGC~5775. From top to bottom, we show vertical profiles of the non-thermal intensities at $140~\rm MHz$ and $1500~\rm MHz$, respectively, and the non-thermal radio spectral index. Solid lines show the best-fitting advection model. The intensities were normalised with respect to the 140-MHz data point at $z=0~\rm kpc$.}
\label{figure:wind_sw}
\end{figure*}

\subsection{Useful analytical expressions}
\label{ass:useful_analytical_expressions}
In this appendix, we present a few analytical expressions that are useful for the analysis in the main text.
\label{as:linearisation_of_the_wind_solution}
\subsubsection{Linearised wind velocity profile}
We show that within the flux tube scale height $z_0$, the wind velocity profile can be well approximated by the following linearised equation:
\begin{equation}
    v = v_0 \left (1+\frac{z}{z_0} \right )
    \label{aeq:linearised_solution}
\end{equation}
Here $v_0$ is the wind velocity at $z=0~\rm kpc$. We now define $\tilde{z}\equiv z/z_0$ and insert this into equation~\eqref{aeq:wind_equation},
\begin{equation}
    (1+\tilde{z})^2 - \log(1+\tilde{z})^2 = 2\log(1+\tilde{z}^2) + 1,
\end{equation}
where we have chosen the integration constant $C$ accordingly. For $\tilde{z}\ll 1$, we can neglect terms of $\tilde{z}^2$,
\begin{equation}
    1 + 2\tilde{z} - \log(1+2\tilde{z}) = 1.
\end{equation}
With $\log(1+\tilde{z})\approx \tilde{z}$,
\begin{equation}
    1 + 2\tilde{z} - 2\tilde{z} = 1.
\end{equation}
This equation is fulfilled everywhere, so that the solution in equation~\eqref{aeq:linearised_solution} is indeed a solution of the wind equation~\eqref{aeq:wind_equation}. We checked that this linearisation is indeed a good approximation of the wind velocity profile as long as $z\lesssim z_0$. Equation~\eqref{aeq:linearised_solution} can also be recast into a slightly different form:
\begin{equation}
    v = v_{\rm c} \left ( 1 + \frac{z-z_{\rm c}}{z_0} \right ),
\end{equation}
meaning the initial wind velocity is $v_0=v_{\rm c}(1-z_{\rm c}/z_0)$.

\subsubsection{Thermal electron density scale height}
\label{ass:thermal_electron_density_scale_height}
Because of the continuity equation~\eqref{aeq:continuity},
\begin{equation}
    \frac{\rho_0}{\rho} = (1+\tilde{z}^2)(1+\tilde {z}),
    \label{aeq:density_continuity}
\end{equation}
where we made use of the flux tube geometry assuming $\beta=2$ and used the linearised velocity profile from equation~\eqref{aeq:linearised_solution}. Here $\rho_0$ is the density in the midplane and again $\tilde z = z/z_0$. Without proof, we state that equation~\eqref{aeq:density_continuity} can be well approximated by:
\begin{equation}
        \frac{\rho_0}{\rho} = \exp[2\log(2)\tilde{z}], 
\end{equation}
in case $\tilde{z}\lesssim 2$. Hence, the vertical profile of the density and therefore of the thermal electron density can be well fitted by an exponential function, so that $n_{\rm e}=n_{\rm e,0}\exp(-z/h_{\rm e})$. The scale height of the thermal electron density is:
\begin{equation}
    h_{\rm e} = \frac{z_0}{2\log{(2)}}.
    \label{aeq:thermal_electron_density_scale_height}
\end{equation}
We found that this indeed a good approximation in all four quadrants. The largest deviation is found in the NW quadrant because $\beta$ deviates most from the assumed value of $\beta=2$.

\begin{figure}
 \centering
    \includegraphics[width=\hsize]{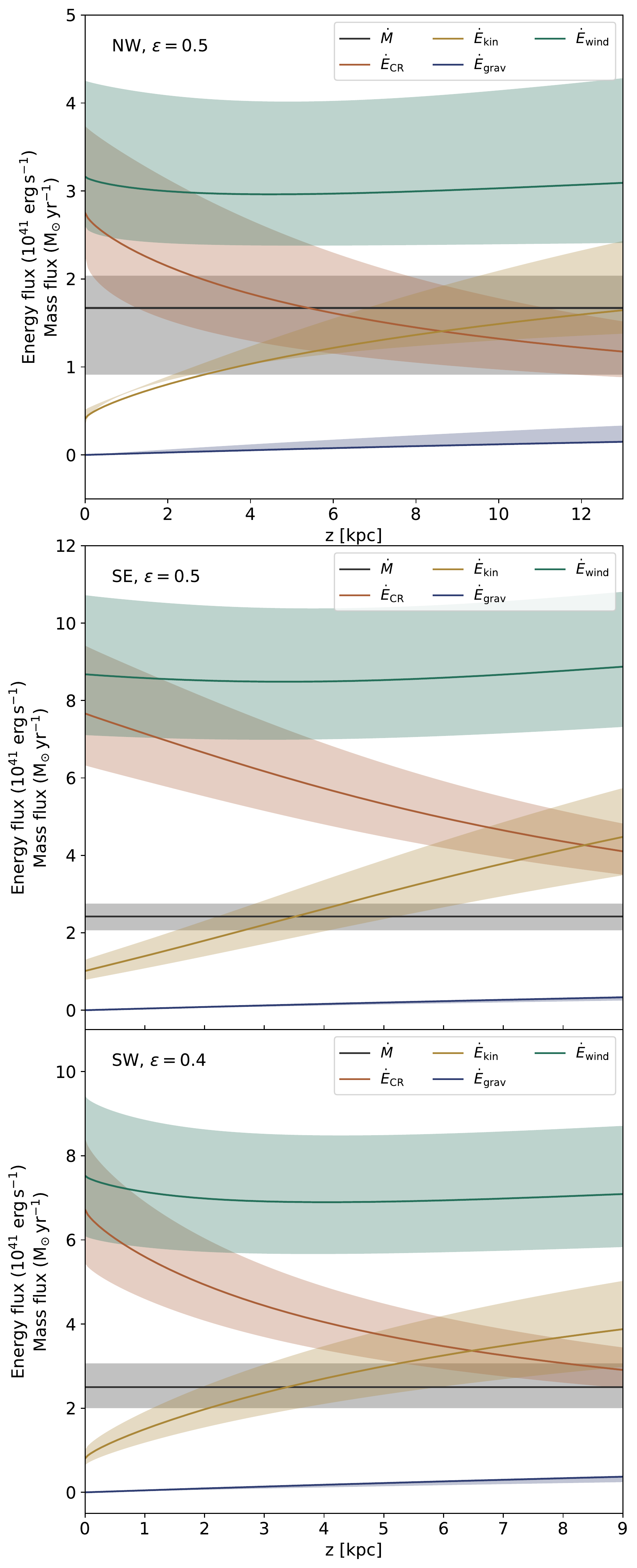}
    \caption{Energy and mass fluxes in the north-western (top), south-eastern (middle), and south-western (bottom) halo (extrapolated to an entire hemisphere) for an entrainment efficiency of $\epsilon = 0.5$ ($\epsilon =0.4$ in the south-western halo). The mass flux $\dot M$ is constant due to the continuity equation. The CR energy flux decreases, the energy of which is transferred into the kinetic energy of the gas and the work lifting the gas in the gravitational potential. The total energy flux of the wind is approximately constant.}
    \label{fig:flux}
\end{figure}

\subsubsection{Vertical gradient of rotation speed}
\label{ass:rotation_velocity_gradient}
We present here a derivation of the rotation velocity gradient as expected for our model. The vertical gradient of the rotation speed is:
\begin{equation}
    \diff{v_{\rm rot}(z)}{z} =  v_{\rm rot} \diff{}{z} \left (  \frac{r_0}{r} \right ),
\end{equation}
where $v_{\rm rot}$ is the maximum rotation speed in the galactic midplane. This equates to:
\begin{equation}
    \diff{v_{\rm rot}(z)}{z} = -v_{\rm rot}\frac{r_0}{r^2} \diff{r}{z}.
    \label{aeq:rotation_gradient}
\end{equation}
The derivation of the outflow radius with height is:
\begin{equation}
\diff{r}{z} = \frac{\tilde{z}}{\sqrt{1+\tilde{z}^2}}\frac{r_0}{z_0},
\end{equation}
where we again used $\tilde{z}=z/z_0$. Neglecting higher terms, assuming $\beta=2$,
and using $\tilde z\approx 1$:
\begin{equation}
    \diff{r}{z} = \frac{r_0}{\sqrt{2}z_0}.
\end{equation}

Plugging this into equation~\eqref{aeq:rotation_gradient}, we can estimate the rotational gradient at $z=z_0$:
\begin{equation}
    \diff{v_{\rm rot}(z)}{z} = -\frac{v_{\rm rot}}{2\sqrt{2} z_0}. 
\end{equation}
Replacing $z_0$ with the the scale height of the thermal electron density using equation~\eqref{aeq:thermal_electron_density_scale_height}, we can express the rotational gradient at $z=z_0$ in this way:
\begin{equation}
    \diff{v_{\rm rot}(z)}{z} = -\frac{v_{\rm rot}}{4\sqrt{2}\log(2) h_{\rm e}}. 
\end{equation}

\subsection{Remaining quadrants}
\label{ass:remaining_quadrants}
Figures \ref{figure:wind_nw}--\ref{figure:wind_sw} show the best-fitting models in the north-western, south-eastern, and south-western quadrants, respectively, and supplement Figure~\ref{figure:wind_ne} in the main text. Figure~\ref{fig:flux} shows the mass and energy fluxes in the remaining quadrants as well.

\section{Cosmic ray-driven wind}
\label{as:cosmic_ray_driven_wind}

In this section, we present a preview of a fully self-consistent wind model using both the cosmic rays and the thermal gas as driving forces \citep[see also the recent paper by][]{2021arXiv210205696Q}. We start from the models of \citet{samui_et_al_2010} and \citet{yu_et_al_2020}, who use a spherical geometry, and adjust this model to a flux tube geometry as in \citet{everett_et_al_2010}. We divide the spatial coordinates into two domains, one with source terms $z\leq z_{\rm sb}$ and one without source terms $z>z_{\rm sb}$ where $z_{\rm sb}$ is the height to which mass and energy are injected into the flow -- i.e.\ the height of the `star burst', for which we assumed $z_{\rm sb}=0.25~\rm kpc$. For a flux tube of cross-section $A$, the equations that need to be solved are then the conservation of mass for $z\leq z_{\rm sb}$:
\begin{equation}
    \frac{1}{A}\frac{\rm d}{{\rm d}z}(\rho v A) = q,
\end{equation}
the conservation of energy for the thermal gas:
\begin{equation}
   \frac{1}{A} \frac{\rm d}{{\rm d}z}\left [ \rho v A\left (\frac{1}{2}v^2 + \frac{\gamma}{\gamma-1}\frac{P_{\rm g}}{\rho}\right )\right ] = -\rho v g + I + Q_{\rm g},
\end{equation}
where the gravitational acceleration $g$ is defined below (Eqn. \ref{eqn:gravacc}), and the conservation of energy for the cosmic-ray gas:
\begin{equation}
\frac{1}{A}\frac{\rm d}{{\rm d}z}\left [ \frac{\gamma_{\rm CR}}{\gamma_{\rm CR}-1}AP_{\rm CR}(v+v_{\rm A})\right ]=-I + Q_{\rm CR}.
\end{equation}
The cosmic-ray energy flux transfer is then:
\begin{equation}
    I=-(v+v_{\rm A})\frac{{\rm d}P_{\rm CR}}{{\rm d}z}.
\end{equation}
Finally, the conservation of momentum translates to:
\begin{equation}
    \rho v \frac{{\rm d}v}{{\rm d}z} = - \frac{{\rm d}P_{\rm g}}{{\rm d}z}- \frac{{\rm d}P_{\rm CR}}{{\rm d}z} -\rho g.
\end{equation}

The source terms for the energy injection by the thermal gas $Q_{\rm g}$ and for the cosmic-ray gas $Q_{\rm CR}$, as well as for the mass injection, $q$, vanish for $z>z_{\rm sb}$ in the above equations. As in \citet{yu_et_al_2020}, we establish the boundary condition as $v=c_{\star}$ at $z=z_{\rm sb}$, where $c_{\star}$ is the compound sound speed as defined by \citet{samui_et_al_2010}. The Alfv{\'e}n speed is modelled as:
\begin{equation}
    v_{\rm A} = \frac{B}{\sqrt{4\upi\rho}},
\end{equation}
where $B=3~\umu\rm G$ is the vertical component of the magnetic field, which we obtained from measurements of the linear polarisation of the radio continuum emission \citep{soida_etal_2011}.
Within the wind, the magnetic field obeys
\begin{equation}
  \frac{\rm d}{{\rm d}z} (BA) = 0.
\end{equation}

Since the composite speed of sound is not constant in this model, the gravitational acceleration needs to decrease near the galactic mid-plane so that the wind can pass the critical point \citep{yu_et_al_2020}. We chose the following parametrisation of the gravitational acceleration:
\begin{eqnarray}\label{eqn:gravacc}
    g(z) & = & \left ( \frac{v_{\rm  rot}^2}{2r_0}\right ) \left (\frac{z}{1~\rm kpc}\right ) \quad (z\leq 1~\rm kpc) \\\nonumber
    & = &  \frac{v_{\rm rot}^2}{2r_0}\exp\left (-\frac{z}{h_{\rm grav}}\right )\quad (z>1~\rm kpc),
\end{eqnarray}
with $h_{\rm grav}=4r_0$ so that the gravitational acceleration is identical to our iso-thermal wind model at $z>1~\rm kpc$.

\begin{figure}
 \centering
    \includegraphics[width=\hsize]{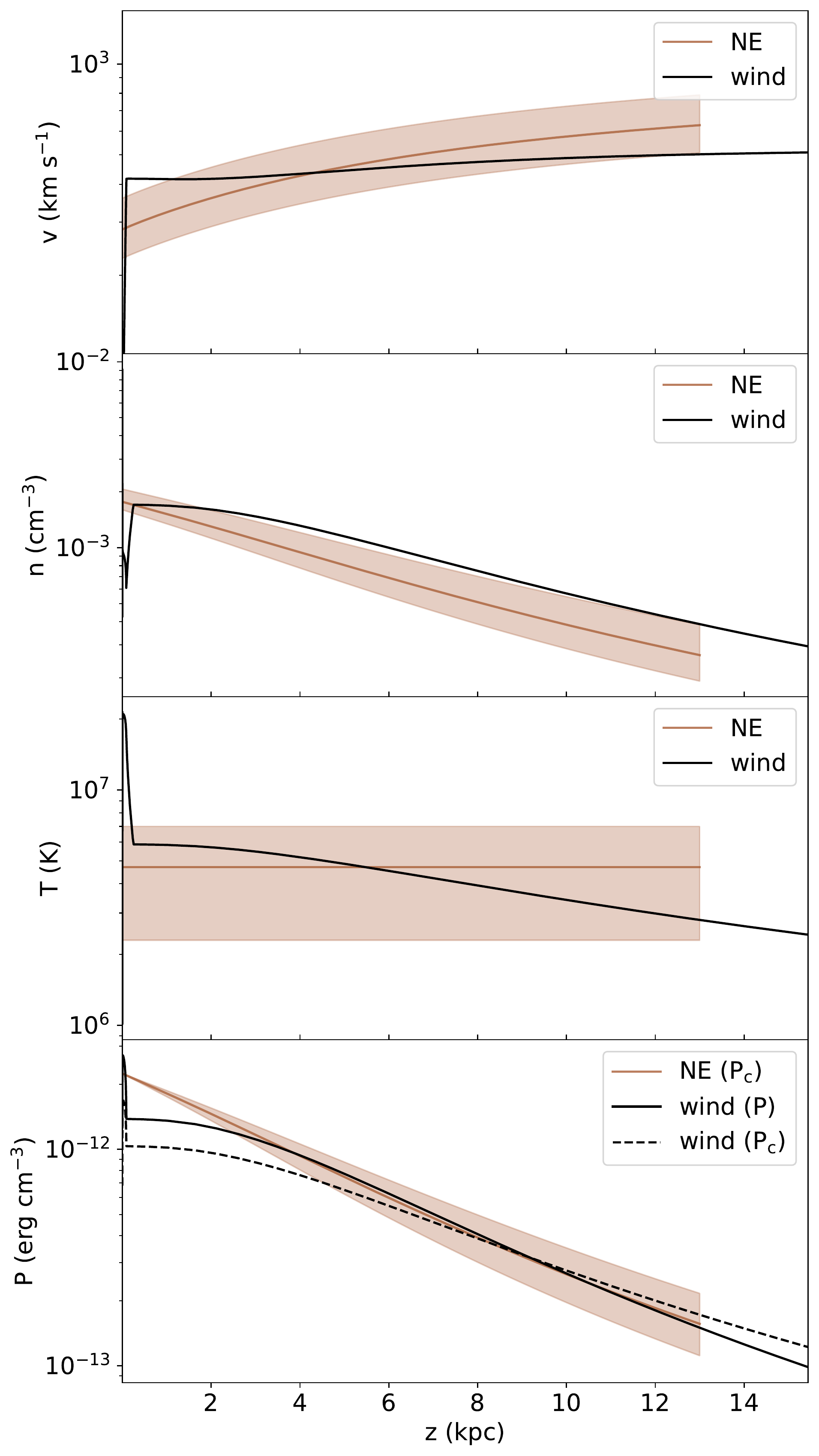}
    \caption{Example wind model using the fully self-consistent equations of conservation of mass, momentum and energy. The parameters were chosen to resemble the wind velocities, densities and pressures typically found in our four quadrants. For comparison we show the best-fitting iso-thermal wind model in the north-eastern quadrant. See text for details.}
    \label{fig:wind_profiles}
\end{figure}

Details of this approach will be presented in a forthcoming paper. Here we only check for consistency of the physical parameters with our iso-thermal wind solution. As in the main text, we chose $r_0=7~\rm kpc$ and a flux tube scale height of $z_0=8~\rm kpc$ in approximate agreement with our four quadrants. The energy injection rate is $\dot E=1.5\times 10^{42}~\rm erg\,s^{-1}$ with half channelled into the thermal energy injection and the other half into the cosmic-ray energy injection. These values are consistent with those discussed in Section~\ref{section:discussion}. As can be seen in Fig.~\ref{fig:wind_profiles}, the resulting vertical profiles of the wind velocity, density and pressure are in good agreement with our results in Section~\ref{section:properties}. An exception is the more modest increase in the wind velocity, which in the self-consistent model only rises from $\approx$400~$\rm km\,s^{-1}$ at $z=z_{\rm sb}$ to $\approx$500~$\rm km\,s^{-1}$. On the other hand, the density decreases from $4\times 10^{-3}~\rm cm^{-3}$ to $4\times 10^{-4}~\rm cm^{-3}$ at the edge of the halo at $z=15~\rm kpc$, and the pressures are also in good agreement with our mid-plane pressures for both the thermal and the cosmic-ray gas (Section~\ref{sss:magnetic_field_strength}) and the decrease of the cosmic-ray gas pressure in the halo. The calculated mass-loss rate is with $\dot M=15~\rm M_{\sun}\,yr^{-1}$ approximately twice as high as our previously calculated rate. The reason is that this model takes the dynamical effect of the thermal gas into account, whereas our mass-loss rate conservatively assumed that the total energy rate is limited by the cosmic ray injection (Section~\ref{section:mass_loss_rate}). In any case, the assumption of an entrainment factor $\epsilon$ of order unity is well supported by this model.

\bsp

\label{lastpage}

\end{document}